\documentclass[prd,onecolumn,amsmath,amssymb,floatfix,superscriptaddress,notitlepage,nofootinbib,preprintnumbers]{revtex4-2}

\usepackage[T1]{fontenc}\usepackage[utf8]{inputenc}\usepackage{lmodern}
\usepackage{amsmath,amssymb,mathtools,bm}
\usepackage[margin=0.95in]{geometry}
\usepackage{xcolor}\usepackage{booktabs,longtable,array}\usepackage{enumitem}
\usepackage{graphicx}
\usepackage{placeins}
\usepackage{hyperref}
\usepackage{natbib}

\newcommand{\R}{\mathcal R}
\newcommand{\Hc}{\mathcal H}
\newcommand{\e}{\epsilon}
\newcommand{\dd}{\mathrm d}
\newcommand{\Order}{\mathcal O}
\newcommand{\threeR}{{}^{(3)}\!R}
\newcommand{\unit}[1]{\hat{#1}}
\newcommand{\Dr}{\Delta\rho}

\newcommand{\Kur}{\mathcal K}
\newcommand{\dK}{\delta_\Kur}
\newcommand{\dKcomp}[1]{\delta_{\Kur,\mathrm{#1}}}
\newcommand{\lapse}{\mathcal A} % this paper's comoving-gauge g_{00} perturbation

\definecolor{srcblue}{rgb}{0.10,0.25,0.60}
\definecolor{todored}{rgb}{0.65,0.10,0.10}
\definecolor{softgray}{rgb}{0.96,0.96,0.96}

\begin{document}

\title{Not ${\cal R}$ Kurvature: Beating Large-Scale White Noise}% 

\author{Wayne Hu}
\affiliation{Kavli Institute for Cosmological Physics, Enrico Fermi Institute, and Department of Astronomy \& Astrophysics, University of Chicago, Chicago IL 60637
}

\begin{abstract}
Kurvature, a recently identified curvature invariant, has been argued to generically acquire superhorizon, or large-scale, white noise from hard-hard momentum mode coupling,  even when only small nonlinearities are present in the matter sector.  If kurvature were related directly to cosmological curvature perturbations ${\cal R}$ via a Poisson equation, this white noise would cause an infrared divergence in its variance that is  observably sensitive to the ultraviolet physics of the hard-hard modes.  However this relation does not generically hold.  
Kurvature is not an intrinsic 3-curvature: on comoving slices, it contains extrinsic-curvature terms, and the intrinsic curvature itself is not related to ${\cal R}$ by a Poisson equation beyond linear order.   We test this Poisson relation directly with second-order perturbation theory in radiation domination, relevant to Cosmic Microwave Background observables.  We verify that quadratic hard-hard composites do indeed cause large-scale white noise in the kurvature density, but 
the Hamiltonian constraint separates these contributions into intrinsic curvature and extrinsic shear or alternately density and extrinsic expansion.   Only the extrinsic terms carry the growing contribution to the dimensionless kurvature density $\dK$ which seemingly mimics ordinary density fluctuations above the horizon.   The hard-hard modes produce negligible and ultraviolet-convergent contributions to the curvature power spectrum, leaving no IR relic in ${\cal R}$ from purely ultraviolet modes, with dominant contributions from the horizon-sized modes at the epoch of evaluation.  
By contrast, the Poisson construction of a conjectured curvature potential from kurvature is indeed infrared divergent and ultraviolet cutoff sensitive. The kurvature white noise employed in that construction arises from the extrinsic curvature associated with acoustic beat modes of a radiation fluid that compresses and shears in its nonlinear evolution. 
 \end{abstract}

\maketitle

\section{Introduction}
\label{sec:intro}

Barenboim, Ireland and Stebbins (BIS) \cite{Barenboim:2025ccc} show that the generation of white noise on
large scales in a certain curvature invariant dubbed kurvature is a generic property of any matter component governed by local nonlinear partial differential equations.  Kurvature is closely related to the matter density and curvature invariants in the center-of-momentum frame of the matter and, in linear theory, is related to the usual cosmological adiabatic curvature mode $\R$ for spatial metric fluctuations through a Poisson-like relation, which we call the BIS relation.   White noise in the former would seemingly imply an infrared-divergent $k_L^{-1}$ contribution to the variance of $\R$, through rescaling its power by $k_L^{-4}$, where $k_L$ is the large-scale wavenumber.   BIS-II \cite{Barenboim:2025jdg} uses this Poisson-like relation to argue that an observable relic of kurvature white noise remains in the Cosmic Microwave Background (CMB) as an ordinary curvature mode, for which the Sachs-Wolfe effect provides a temperature anisotropy of $\Delta T/T \approx {\cal R}/5$, and constrains the UV physics of the matter sector, specifically by requiring a cutoff or reduction in the initial power spectrum of the high-momentum hard modes that generate it.   

While large-scale white noise in the kurvature density may be generic, its posited relic in ordinary curvature modes is not.  BIS-II conjectures that inevitable early generation of kurvature white noise in the radiation-dominated universe produces this UV-sensitive relic, and BIS explicitly give the nonlinear evolution of acoustic waves in an irrotational perfect fluid as an example mechanism.   Here we directly examine whether the BIS-II conjecture holds for this system in second-order perturbation theory and clarify the difference between kurvature, intrinsic curvature, and ordinary curvature modes.  

The remainder of this paper is organized as follows.  Section~\ref{sec:kurvature} defines kurvature in the matter center-of-momentum frame, uses the Hamiltonian constraint to separate its intrinsic and extrinsic contributions, and identifies the linear BIS relation whose nonlinear extension is at issue.  Section~\ref{sec:radiation} then gives an explicit second-order calculation for an irrotational radiation fluid in comoving gauge and directly shows that while the kurvature does acquire large-scale white noise (LSWN) it does so from the extrinsic curvature, while the curvature mode does not acquire an IR-divergent term. 
 Section~\ref{sec:disc} discusses the physical interpretation of these results.   In a series of Appendices, we provide further checks and physical insight on the remaining hard-hard LSWN.  Appendix~\ref{app:dewitt} develops the extrinsic-curvature interpretation and directly evaluates the expansion contribution to kurvature; Appendix~\ref{app:multipoles} details the curvature and density multipoles and evolution; Appendix~\ref{app:synch} gives an independent synchronous-gauge construction and pullback to comoving slicing and threading; and Appendix~\ref{app:lv} reconciles the comoving curvature result with the exact Langlois--Vernizzi curvature conservation law \cite{Langlois:2005qp}.

\section{Kurvature}
\label{sec:kurvature}

Following
Stebbins~\cite{Stebbins:2023gic,Stebbins:2026stf}, we define the kurvature through a covariant construction and associate it with coordinate-dependent quantities in the center of momentum (CM) frame.  
With $u^\alpha$ as the unit normalized future-directed timelike eigenvector of the total stress tensor,
$T^\alpha{}_{\beta}u^\beta=-\rho\,u^\alpha$, or, equivalently by the Einstein
equation, $G^\alpha{}_{\beta}u^\beta=-8\pi G\rho\,u^\alpha$,  where $\rho$ is the
energy density in the center-of-momentum (CM) frame.  As we shall see below, for a fluid $u^\alpha$ is the fluid 4-velocity.  Defining 
$\theta\equiv u^\alpha{}_{;\alpha}$ as its local expansion rate, he calls
\begin{equation}
 \Kur\equiv\frac{8\pi G\rho}{3}-\frac19\theta^2 ,
 \qquad
 \Dr\equiv\frac{3\Kur}{8\pi G}=\rho-\frac{1}{24\pi G}\,\theta^2,
 \label{eq:kurvature}
\end{equation}
the \emph{kurvature} and kurvature density, respectively; BIS adopt this
construction and develop its implications for LSWN.
We write the kurvature $\Kur$ in script, reserving plain $K$ for the trace
of the extrinsic curvature.  On the preferred slices picked out by $u^\alpha$ and used below for an irrotational fluid,
$K^2=\theta^2$, so $\Kur=8\pi G\rho/3-K^2/9$.   Nonetheless, both $\Kur$ and $\Dr$ are
built from the local Einstein tensor and the CM four-velocity, so both are covariant
and locally measurable.  The Stebbins spelling of kurvature is deliberate, to keep the quantity distinct from the
several other curvatures in use.  When specializing to an irrotational radiation fluid below it will be convenient to work with  the dimensionless,
background-normalized kurvature density, $\dK\equiv\Dr/\bar\rho$ (Eq.~\eqref{eq:allocdK}), the same
construction as the ordinary density contrast $\delta\equiv\delta\rho/\bar\rho$ but with very different interpretations.

This distinction makes an essential point:  $\Dr$ is the energy
density \emph{with the squared local expansion rate, $\theta^2/24\pi G$, subtracted}.
It is a curvature, not
$\delta\rho$, and so naive objections to white noise fluctuations based on causality and conservation need not apply.  
Instead, the leading nonlinear source for $\Dr$ is quadratic in the kinematics of the CM
congruence: its expansion $\theta$, trace-free shear $\sigma_{\alpha\beta}$, and,
for a general flow, vorticity $\omega_{\alpha\beta}$, which measures local rotation.
The Raychaudhuri equation makes their roles physical: shear focuses neighboring CM
worldlines, whereas rotation opposes that focusing.  They consequently enter the
general source as $\sigma^2-\omega^2$ \cite{Stebbins:2026stf}, with no spatial derivatives, and can generate
a  LSWN floor from hard-hard beat coupling of modes.

The BIS-II association of LSWN kurvature with an observable IR divergence from a UV-sensitive cutoff in hard modes involves its early generation in the radiation-dominated universe from the usual adiabatic scalar fluctuations from inflation.   To cleanly evaluate this claim, we use a scalar, irrotational,
perfect-fluid branch, ignoring the viscosity and heat conduction of the actual photon-baryon fluid, which provide the familiar UV cutoff through diffusion damping, as well as vorticity and vector modes, which lie beyond the scalar sector treated here.
With vorticity absent, the CM flow is normal to the comoving
slices.  

In Sec.~\ref{sec:ham}, we show how the  Hamiltonian constraint in this system usefully partitions  LSWN kurvature into intrinsic
three-curvature and extrinsic shear pieces rather than energy density $\rho$ and expansion $\theta^2$.  In Sec.~\ref{sec:bridge}, we show that beyond linear theory,  the intrinsic piece is not simply related to the curvature potential via a Poisson-like relation, which we call the BIS relation.

\subsection{Hamiltonian constraint}
\label{sec:ham}

Now let us move to the preferred slicing of the kurvature construction.
First recall the general Arnowitt-Deser-Misner (ADM) construction.
Foliate spacetime into spacelike slices with unit normal $n^\mu$, induced metric
$\gamma_{ij}$, and extrinsic curvature
$K_{ij}$, split into trace and trace-free
parts.  The normal--normal projection of the Einstein
 equations is given by \begin{equation}
2G_{\mu\nu}n^\mu n^\nu= \threeR+K^2-K_{ij}K^{ij}=16\pi G\,E_n ,
 \qquad E_n\equiv T_{\mu\nu}n^\mu n^\nu ,
 \label{eq:ham}
\end{equation}
which we call the Hamiltonian constraint. 

The preferred slicing for our perfect fluid system is the comoving one, $n^\mu=u^\mu$,  where $u^\mu$ is the fluid 4-velocity.   This construction applies  only if the matter
is irrotational --- by the  Frobenius theorem, a unit timelike congruence is hypersurface-orthogonal
precisely when its vorticity vanishes.  For this system the 4-velocity coincides with the BIS definition of $u^\mu$ in Eq.~\eqref{eq:kurvature}: for
$T^\mu{}_\nu=(\rho+P)u^\mu u_\nu+P\delta^\mu{}_\nu$,  the eigenvector condition $T^\mu{}_{\nu}u^\nu=-\rho\,u^\mu$ holds, and $E_n=\rho$.   
On the comoving slice $K_{ij}=-\tfrac13\theta\gamma_{ij}-\sigma_{ij}$, so $K=-\theta$
and $K_{ij}K^{ij}=\theta^2/3+2\sigma^2$ with $\sigma^2\equiv\tfrac12\sigma_{ij}\sigma^{ij}$. Eq.~\eqref{eq:ham} becomes
$\threeR=16\pi G\rho-\tfrac23\theta^2+2\sigma^2$, and eliminating $\theta^2$ in favor
of $\Kur$ via Eq.~\eqref{eq:kurvature} gives the exact relation for kurvature \cite{Tsagas:2007yx,Stebbins:2023gic}
\begin{equation}
  \Kur=\frac{\threeR}{6}-\frac{\sigma^2}{3}
   \label{eq:gaussexact}
\end{equation}
in terms of intrinsic curvature and shear in the extrinsic curvature.  With this partitioning of kurvature, the central question is whether LSWN in $\Kur$ comes from intrinsic curvature or shear or equivalently whether the RHS of 
\begin{equation}
 \threeR=16\pi G\,\Dr+2\sigma^2
\end{equation}
predominantly cancels or gives $\threeR = 16\pi G\times \Order(\Dr)$.
Furthermore,  BIS-II  seeks to equate $\Delta\rho$ through $\threeR$ to observable curvature modes $\R$ by inverting a Poisson-like relation.
This is a correct relation in linear theory where the shear contribution $\sigma^2$ is second order and the 3-curvature and $\R$ are related by a Poisson-like equation
\begin{equation}
\nabla^2\R^{(1)} =-\frac{1}{4} a^2 \delta\threeR^{(1)} = -4\pi G a^2\,\Dr^{(1)} , \qquad (\text{BIS relation}).
\label{eq:BIS-linear}
\end{equation}
Here and throughout superscripted parentheses following variables denote the order of the perturbative expansion.

Deferring its status beyond linear theory to the next section,
first let us establish what kind of behavior in $\Dr$ might behave as a purported ordinary constant curvature relic above the horizon with it.   For an ordinary first-order curvature mode we can associate
$\delta \threeR=-4a^{-2}\nabla^2\R$ with $\R$ frozen on super-horizon scales.  Thus 
$\threeR\propto a^{-2}$, a constant 3-curvature fluctuation in comoving coordinates.  Therefore by Eq.~\eqref{eq:gaussexact} with
$\sigma^2$ set aside, this requires $\Dr\propto a^{-2}$ as well.  In the radiation era, where
$\bar\rho\propto a^{-4}$, this is the statement $\dK\equiv\Dr/\bar\rho\propto a^2$: a
$\dK$ that grows at the same rate as the comoving density perturbation $\delta$ above the horizon, rather than a constant or decaying dimensionless kurvature density.   
We shall see by explicit calculation, while this does occur for the radiation fluid, it comes exclusively from $\sigma^2$.

\subsection{BIS-II conjecture and second-order curvature}
\label{sec:bridge}

The form of the BIS relation in Eq.~\eqref{eq:BIS-linear} is exact in linear theory, but the kurvature density beyond linear theory involves the extrinsic curvature. Expanding Eq.~\eqref{eq:gaussexact} to second order in the perturbations gives
\begin{equation}
 \Dr_{(2)}={\frac{\delta\threeR^{(2)}\big|_{\rm com}}{16\pi G}}
 \;-\;{\frac{(\sigma^{(1)})^2}{8\pi G}}.
 \label{eq:gauss2}
\end{equation}
Notice that there is no second-order density or quadratic expansion contribution in this form: it is composed purely of second-order intrinsic curvature and hard-hard first-order shear terms.  This already establishes the LSWN kurvature mechanism at second order: the composite first-order shear modes of high wavenumber beat couple to a low wavenumber in their quadratic combination.  Shear modes from the acoustically compressing radiation fluid are expected but do not on their own imply LSWN in the intrinsic curvature.

Furthermore at second order, the usual curvature mode $\R$ is not related to $\delta\threeR$ through a simple Poisson-like equation.  In the isotropic threading we use below, $\gamma_{ij}=a^2(1+2\R)\delta_{ij}$,
and the construction for a three-dimensional Ricci
scalar gives, to the order needed here,
\begin{equation}
 a^2\delta\threeR^{(2)}=-4\nabla^2\R^{(2)}
 +16\,\R^{(1)}\nabla^2\R^{(1)}+6\big(\partial_i\R^{(1)}\big)^2 .
 \label{eq:R3geometric}
\end{equation}
The first term is linear in the second-order curvature perturbation itself. The second and third terms are built entirely from products of
the \emph{first-order} field.   Thus beat coupling of the hard modes also generically implies some level of LSWN in the intrinsic curvature but not of the growing type in $\dK$.  An important thing to note here is that beyond linear theory the ordinary curvature mode $\R$ is not simply the  potential for 3-curvature fluctuations.

The BIS-II inference involves two logically distinct steps.  At the epoch when hard modes generate kurvature, it identifies the hard-hard density response with an accompanying  intrinsic-curvature perturbation,
\begin{equation}
 -\frac14a^2\delta\threeR^{(2)}\big|_{\rm com}\approx -4\pi G\,a^2\,\Dr^{(2)}, \qquad (\text{BIS-II conjecture}).
\label{eq:BIS-extrapolation}
\end{equation}
It then assumes that this inferred curvature perturbation is a relic regular adiabatic mode and evolves it in linear theory, where the BIS relation holds.  For the BIS-II construction plotted below, this identification defines the inferred hard-hard curvature response
\begin{equation}
\nabla^2\R_{\rm BIS} \equiv -4\pi G a^2\,\Dr^{(2)} .
\label{eq:BIS-R}
\end{equation}
The question addressed here is whether the BIS-II conjecture itself holds: whether the nonlinear hard-hard response supplies the intrinsic-curvature initial data required for that linear interpretation.  Kurvature is not 3-curvature, and a Poisson potential $\R_{\rm BIS}$ constructed from kurvature as in Eq.~\eqref{eq:BIS-R}  is not thereby an ordinary curvature mode.

We next turn to the explicit calculation of kurvature and its correspondence with $\R$ for a radiation fluid in second-order perturbation theory to illustrate these distinctions.

\section{Radiation LSWN}
\label{sec:radiation}
In this section we carry out the direct calculation in radiation domination to second order.  We first fix the comoving-gauge conventions and linear acoustic solutions, then construct the quadratic hard-hard sources and solve the coupled constraint and evolution equations for the soft LSWN mode.  We distinguish between the second order curvature response $\R$ and that of the kurvature density fluctuation $\dK$.  For $\dK$ we identify which intrinsic and extrinsic geometric terms supply the second order source and  directly test whether they generate the inverse-Laplacian curvature response assumed in the BIS-II conjecture and the associated Eq.~\eqref{eq:BIS-R}.  We then assemble the corresponding contribution to the power spectrum of the hard-hard modes alone, determine which hard modes are the dominant contributors, and contrast this finite direct result with the cutoff-sensitive spectrum obtained from the BIS-II conjecture.

\subsection{Notation conventions}

We work in the radiation era with conformal time $\eta$, $a\propto\eta$, $\Hc\equiv \dot a/a= 1/\eta$,
$w=c_s^2=1/3$, and write the line element with scalar metric perturbations 
\begin{equation}
 \frac{\dd s^2}{a^2}=-(1+2\lapse)\,\dd\eta^2+2\partial_iB\,\dd\eta\,\dd x^i
 +(1+2\R)\,\delta_{ij}\,\dd x^i\dd x^j .
 \label{eq:metric}
\end{equation}
Throughout, Latin indices are spatial.  In the perturbative and Fourier algebra they
are raised, lowered, and contracted with the background $\delta_{ij}$; repeated
spatial indices are summed regardless of placement, e.g.\
$v_iv_i\equiv\delta^{ij}v_iv_j$.
The gauge is scalar-velocity orthogonal in slicing and isotropically threaded, i.e.\ the preferred slicing is orthogonal to the scalar component of the fluid 4-velocity since we are interested in curvature generation and the scalar sector.   There is a small transverse velocity component that represents the vector sector which is not treated here.
 Note that
``comoving'' does not mean the shift vanishes: the scalar coordinate velocity
potential is $v=-B$.  Nonetheless we call this gauge ``comoving'' as shorthand.

We are interested in hard-hard quadratic combinations that beat couple to source soft momenta modes that can exhibit LSWN.  We refer to these hard modes as the $q$ and $p$ legs that close the triangle with a long wavelength $k_L$ mode:
\begin{equation}
 \begin{aligned}
 \bm p&=\bm k_L-\bm q, &
 \e&\equiv\frac{k_L}{q}, &
 \mu&\equiv\unit {\bm k}_L\cdot\unit {\bm q},
 \end{aligned}
 \end{equation}
 which then implies
 \begin{equation}
 \begin{aligned}
 \bm k_L\cdot\bm q&=\e\mu q^2, &
 \bm q\cdot\bm p&=q^2(\e\mu-1), &
 \frac pq&=\sqrt{1-2\e\mu+\e^2} .
 \end{aligned}
 \label{eq:geometry}
\end{equation}
Thus we will generally view $\bm q$ as the principal hard mode and express all other quantities in terms of $\epsilon$, which characterizes how squeezed the triangles are, and $\mu$ which jointly determines the triangle shape and implicitly $\bm p$.
For compactness of the expressions it will be convenient occasionally to rescale the time variable $\eta$ to the dimensionless acoustic time for the hard mode $q$:  
$x\equiv c_s q\eta= q\eta/\sqrt3$.  When it is important to keep the temporal dependence of both modes we explicitly write $x_p\equiv p\eta/\sqrt3=(p/q)\,x$ but generally use $x$ throughout.  Note also that
$\sqrt3x=q/\Hc$.  The limit of interest is
$\e\to0$ at fixed $x$.  At late times $\e^{-1}\gg x\gg 1$, this means the hard modes are acoustically oscillating while the soft mode is outside the horizon $k_L \eta\ll 1$ appropriate for the LSWN calculation.  

Quadratic composites of the $q$ and $p$ modes are defined by the physical unequal-leg coefficient that multiplies the initial curvature mode $\R_i$, whose power spectrum is given by
\begin{equation}
 \big\langle\R_i(\bm k)\R_i(\bm k')\big\rangle
 =(2\pi)^3\delta_{\rm D}(\bm k+\bm k')P_{\R_i}(k),
 \qquad
 \Delta^2_\R(k)=\frac{k^3P_\R(k)}{2\pi^2} = A_s \left( \frac{k}{k_0}\right)^{n_s-1}.
 \label{eq:conventions}
\end{equation}
 With two
independent hard legs of amplitude $A$ and $B$ carrying primordial curvature
$\R_i(\bm q)$ and $\R_i(\bm p)$,
\begin{equation}
 F=F_0+A\,F_q+B\,F_p+AB\,F_{qp}+\Order(A^2,B^2),
 \label{eq:ABconv}
\end{equation}
and every quantity written with a subscript 2 below means $F_{qp}$.  Note that in this ``AB'' form, the $q-p$ permutation double counting is taken care of in the sum over all modes rather than in the $F_{qp}$ kernel itself.

\subsection{First-order solutions}
\label{sec:first}

The usual first-order solutions become the sources for the second-order evolution.   
In general, of the 4 Einstein equations and 2 scalar matter conservation equations, we need only keep 4 independent equations due to the Bianchi identities.  In comoving gauge it is advantageous to take these as the $00$ Hamiltonian equation, the $0i$ momentum equation, the trace-free $ij$ equation, and the matter Euler equation rather than also take the spatial trace of the Einstein equations.   For the principal hard-mode momentum $q$, these give
\begin{align}
 -6\Hc\dot\R_1-2q^2\R_1+6\Hc^2\lapse_1-2\Hc q^2B_1+3\Hc^2\delta_1&=0 ,
 \label{eq:linH}\\
 \dot\R_1-\Hc\lapse_1&=0 ,
 \label{eq:linM}\\
 \lapse_1+\dot B_1+2\Hc B_1+\R_1&=0 ,
 \label{eq:linT}\\
 \lapse_1 +\tfrac14\delta_1&=0 .
 \label{eq:linE}
\end{align}
  Eliminating the constraints,
\begin{equation}
 \delta_1=\frac{2q^2(\R_1+\Hc B_1)}{3\Hc^2},\qquad
 B_1=-\frac{\R_1}{\Hc}-\frac{6\dot\R_1}{q^2},\qquad \lapse_1=\frac{\dot\R_1}{\Hc},
\end{equation}
and substituting into the trace-free Eq.~\eqref{eq:linT} leaves a single propagating scalar, 
\begin{equation}
 \ddot\R_1+2\Hc\dot\R_1+\tfrac13q^2\R_1=0
\end{equation}
with first-order transfer functions $\R^{(1)}(\bm q,x) = \R_1(x) \R_i(\bm q)$,  
$\lapse^{(1)}(\bm q,x) = \lapse_1(x) \R_i(\bm q)$, etc as\begin{equation}
 \R_1=\frac{\sin x}{x},\qquad
 \lapse_1=\cos x-\frac{\sin x}{x},\qquad
 \delta_1=-4\lapse_1,
 \label{eq:transfers}
\end{equation}
\begin{equation}
 b\equiv qB_1=\sqrt3\left(-\sin x-\frac{2\cos x}{x}+\frac{2\sin x}{x^2}\right).
 \label{eq:transfersb}
\end{equation}
The transfer functions also fix the expansion $K=-\theta$ and shear as
\begin{equation}
 \begin{aligned}
 \delta\theta_1&=\frac{1}{a}\left[3\bigl(\dot\R_1 -\Hc\lapse_1 \bigr)+q^2B_1\right]
 =\frac{qb(x)}{a},\\
 \sigma_1{}^i {}_j&=\frac{qb(x)}{a}\left(\hat q^i\hat q_j-\tfrac13\delta^i_j\right),
 \end{aligned}
 \label{eq:sigma1}
\end{equation}
Nothing here is specific to $q$: Eqs.~\eqref{eq:linH}--
\eqref{eq:transfersb} take the same form for the $p$-leg  with $q\to p$, $x\to x_p\equiv(p/q)x$.

\subsection{Second-order sources}
\label{sec:sources}

Quadratic combinations of the $q$ and $p$ first-order solutions become the second-order sources for the soft mode $k_L$.   
Following the standard order-by-order construction (see Ref.~\cite{Inomata:2023faq} for an elaboration in synchronous gauge), we move all products of first-order fields to the right-hand side of the defining equations, where they act as sources for the second order correction to the soft mode.
The  defining equations are the same as in the first order calculation:  3 Einstein equations and the fluid Euler equation.  Following this iterative procedure we obtain
\begin{align}
 -6\Hc\dot\R_2-2k_L^2\R_2+6\Hc^2\lapse_2-2\Hc k_L^2B_2+3\Hc^2\delta_2&=q^2S_H ,\\
 \dot\R_2-\Hc\lapse_2&=qS_M ,\\
 \lapse_2+\dot B_2+2\Hc B_2+\R_2&=S_T ,\\
 \lapse_2+\tfrac14\delta_2&=S_E .
 \label{eq:second}
\end{align}
With $E^\mu{}_\nu=a^2(G^\mu{}_\nu-8\pi G\,T^\mu{}_\nu)$ and $C_\nu\equiv\nabla_\mu T^\mu{}_\nu$ evaluated with only
first-order fields present, the 4 sources are given by
\begin{align}
 \label{eq:proj}
 S_H={}&-\frac{E^0{}_0}{q^2},\qquad
 S_M=-\frac{k_{Li}E^0{}_i}{2iqk_L^2},\qquad
 S_T=-\frac{P^{\rm S}_{ij}E^i{}_j}{k_L^2},\qquad\\
  S_E={}&-\frac{1}{\bar\rho+\bar p}\,\frac{k_{Li}C_i}{ik_L^2}\bigg|_{AB} .
 \label{eq:SEproj}
\end{align}
with $P^{\rm S}_{ij}=\tfrac32(k_{Li}k_{Lj}/k_L^2-\tfrac13\delta_{ij})$.  A fixed
hard-hard pair supplies a direction in addition to $\bm k_L$, so its quadratic $S_T$ residual involves the angles in Eq.~\eqref{eq:geometry}.  Note that we have defined the various second-order sources to be dimensionless by inserting the $q$-scalings on the RHS. 

The $S_E$ source may be less familiar in second-order perturbation theory since it comes from iterating the Euler equation.   Since $C_\nu\equiv\nabla_\mu T^\mu{}_\nu=0$ nonlinearly, its residual is built from first-order metric fluctuations implicit in the covariant derivative and the fluid velocity or equivalently its potential $v_1=-B_1$. 
  Evaluating gives
\begin{equation}
 S_E=%6\lapse_{1q}\lapse_{1p}-\partial_iB_{1q}\,\partial_iB_{1p}=
 6\lapse_{1q}\lapse_{1p}+(\bm q\cdot\bm p)B_{1q}B_{1p},
 \label{eq:SE}
\end{equation}
where we have employed the AB counting procedure.   Another route to the same answer uses Tolman-type reasoning: in comoving gauge, where $v=-B$, the fluid is in slice-by-slice hydrostatic balance, with pressure support counteracting the lapse gradient nonlinearly, or order by order.   The fluid's 4-acceleration is the normal
congruence's, and for a unit normal to a slice of constant coordinate time this is
exactly a gradient of the log-lapse, $a_i=\partial_i\ln N$.
The relativistic Euler equation
$(\rho+p)a_i=-\partial_ip$ and $p=\rho/3$ then integrates to
\begin{equation}
 \ln\frac{N}{a}+\frac14\ln\frac{\rho}{\bar\rho}=f(\eta) ,
 \label{eq:tolman}
\end{equation}
with $f$ spatially homogeneous.  The lapse $N$ itself may be expanded to second order as
\begin{equation}
 \frac{N^2}{a^2}=1+2\lapse+\partial_iB\,\partial_iB+\Order(3) .
 \label{eq:Nshift}
\end{equation}
Extracting the $AB$ coefficient of Eq.~\eqref{eq:tolman} with the first-order relation $\delta_a=-4\lapse_a$ (Eq.~\eqref{eq:linE} for each leg)
 gives
Eq.~\eqref{eq:SE}.

Substituting the first-order solutions into the definitions of the sources, we obtain
\begin{align}
 x^6S_H&=s^2\!\left(-18x^4+42x^2-24\right)+sc\left(4x^5-28x^3+48x\right)
 +10x^4-24x^2 ,\\
 x^4\e^2S_T&=s^2\!\left[-2x^4+12x^2-8+\mu^2\!\left(6x^4-36x^2+24\right)\right]
 \nonumber\\&\quad
 +sc\left[-6x^3+16x+\mu^2\!\left(18x^3-48x\right)\right]
 +x^4-8x^2+\mu^2\!\left(-3x^4+24x^2\right),\\
 \sqrt3\,x^5S_M&=s^2\!\left[13x^4-35x^2+12+\mu^2\!\left(-12x^4+24x^2\right)\right]
 \nonumber\\&\quad
 +sc\left[-3x^5+30x^3-24x+\mu^2\!\left(3x^5-24x^3\right)\right]-7x^4+12x^2 ,\label{eq:SM}\\
 x^4S_E&=s^2\!\left(-9x^4+30x^2-12\right)+sc\left(-24x^3+24x\right)+6x^4-12x^2 ,
 \label{eq:sourcesclosed}
\end{align}
which for $x\ll 1$ takes the limiting forms
 \begin{equation}
 S_H\to-4, \quad \e^2S_T\to-1+3\mu^2,\quad
S_M\to-x(1+\mu^2)/\sqrt3,\quad S_E\to-x^2/3. 
\end{equation}

\subsection{Curvature master equation}
\label{sec:mechanism}

We can now conduct the same steps as in linear theory to eliminate the constraints for $\lapse_2,\delta_2,B_2$ at second order and obtain the sourced master equation for the second-order curvature $\R_2$.  As with the first-order solution this becomes a sourced wave equation rather than a Poisson-like relation of the BIS-II conjecture.
 The momentum equation gives the lapse directly in
terms of $\dot\R_2$ and substituting it into the Euler constraint gives the density:
\begin{equation}
\lapse_2=\frac{\dot\R_2}{\Hc}-\sqrt3x\,S_M,\qquad
 \delta_2=4S_E-\frac{4\dot\R_2}{\Hc}+4\sqrt3x\,S_M .
 \label{eq:recon}
\end{equation}
  Substituting Eq.~\eqref{eq:recon} into the Hamiltonian constraint and
solving for $B_2$ gives
\begin{equation}
 B_2=-\frac{\R_2}{\Hc}-\frac{6\dot\R_2}{k_L^2}+\frac{\Hc\,Q}{2k_L^2},\qquad
 Q\equiv-3x^2S_H+6\sqrt3x\,S_M+12S_E .
 \label{eq:Bsplit}
\end{equation}
 The $B_2$ solution does take a Poisson-like form,
 $B_2\propto k_L^{-2}$ and might seem to admit a BIS-type IR divergence: $Q$ is a
fixed quadratic quantity built from beat coupling of hard-hard modes  and is generically
nonzero as $k_L\to0$.   However $B_2$ is a shift \emph{potential}, not the ADM shift itself.    The ADM shift in the metric is its gradient,
$\beta_i=a^2\partial_iB_2\propto k_L^{-1}$, so for a finite hard convolution its
variance per logarithmic soft interval scales as
$k_L^3|k_LB_2|^2\propto k_L$, so its power spectrum is IR integrable.

Substituting Eqs.~\eqref{eq:recon}--\eqref{eq:Bsplit} into the one equation not yet
used, the trace-free (anisotropy) equation, gives the
master equation,
\begin{equation}
 \ddot\R_2+2\Hc\dot\R_2+\tfrac13k_L^2\R_2=q^2S_R,\qquad
 S_R=-\frac16\e^2S_T-\frac{\sqrt3x}{6}\e^2S_M
 +\frac1{36x^2}\left(Q+xQ'\right) ,
 \label{eq:master}
\end{equation}
with $Q'\equiv dQ/dx$ and $\dot\Hc=-\Hc^2$ in radiation domination.
Explicitly 
\begin{equation}
 \frac1{12}\left(Q+xQ'\right)
 =S_E+xS_E'-\frac34x^2S_H-\frac14x^3S_H'+\sqrt3x\,S_M+\frac{\sqrt3}2x^2S_M' .
 \label{eq:bracket}
\end{equation}
This is a pure function
of $x$ and $\mu$, with no inverse $\e$ or $k_L$ behavior.
Notice that the $\ddot \R_2$ term comes from  $\dot B_2$ and so the naively dangerous $k_L^{-2}$ term in $B_2$ comes in the combination $(6\dot\R_2+\Hc Q/2)k_L^{-2}$ and is removed by multiplying through by $k_L^2$.  The $k_L^{2} \R$ term from the original Hamiltonian constraint does then reappear in Eq.~(\ref{eq:recon}), but not in the form of a Laplacian constraint on $\R$.

This dynamical content then is quite different in form from the BIS-II conjecture with the association in Eq.~\eqref{eq:BIS-R}, which takes the form of a Poisson-like constraint given a source.   Instead,  the second-order hard-hard terms source an acoustic propagation equation~\eqref{eq:master} and produce regular solutions as $k_L \to 0$, or equivalently $\e \to 0$.
 Note that the combined source $S_R$ is itself regular in this limit.
 This is a source-level Laurent test: expanding $S_R$ at fixed $x,\mu$
has no $\e^{-2}$ or $\e^{-1}$ coefficient.
 In particular the  leading order $\e^0$ coefficient of this dimensionless source is 
 \begin{equation}
 S_{R,0}(x,\mu)\equiv\lim_{\e\to0}S_R(x,\e,\mu)
 =\lim_{\e\to0}\left[-\frac16\e^2S_T
 +\frac1{36x^2}\left(Q+xQ'\right)\right]
 \label{eq:SR0assembled}
\end{equation}
and can only generate LSWN $k_L^0$ behavior in $\R_2$ not in $k_L^2 \R_2$ as its use in Eq.~\eqref{eq:BIS-R}  would imply.  

To see this, let us assemble Eq.~\eqref{eq:SR0assembled} from the closed forms of
Sec.~\ref{sec:sources} (Eq.~\eqref{eq:sourcesclosed}),
\begin{equation}
 \begin{aligned}
 3x^4S_{R,0}(x,\mu)=\;&\left(1-12\mu^2\right)+\left(2-6\mu^2\right)x^2\\
 &+\left[\left(12\mu^2-1\right)-18\mu^2x^2-3\left(1-\mu^2\right)x^4\right]\cos2x\\
 &+\left[\left(24\mu^2-2\right)x+\left(1-9\mu^2\right)x^3\right]\sin2x.
 \end{aligned}
 \label{eq:SR0}
\end{equation}
Despite the $x^4$ on the LHS, this source is also regular as $x\to 0$ due to cancellations on the RHS in the limit:
\begin{equation}
 S_{R,0}=\left(\frac13-\mu^2\right)+\Order(x^2).
 \label{eq:SRearly}
\end{equation}
Tracked back to the quadratic form for the first-order solutions of 
Eqs.~\eqref{eq:transfers}--\eqref{eq:transfersb} this comes from $\R_1^2$ as one might expect from a primordial curvature mode on superhorizon scales $x\ll 1$.
With this simple second-order source for $\R_2$, we can proceed next to solve its sourced wave equation.

\subsection{Second-order curvature and kurvature}
\label{sec:Adivergence}

Next let us solve the master equation \eqref{eq:master} in integral form. Its wave nature can be seen by using  $x$ directly as the time variable and rewriting $U(x)\equiv x\R_2(x)$.  Equation~\eqref{eq:master} then becomes
\begin{equation}
 U''+\e^2U=3xS_R,
 \qquad {}'\equiv d/dx.
 \label{eq:Umaster}
\end{equation}
With no independently imposed homogeneous contribution to $U$, its solution is
\begin{equation}
 \R_2(x,\e,\mu)=\frac{3}{\e x}\int_0^x\dd x'\,x'
 \sin\!\big[\e(x-x')\big]\;S_R(x',\e,\mu).
 \end{equation}
 The Laurent expansion again gives no $\e^{-2}$ or $\e^{-1}$, leaving the leading-order LSWN term as 
 \begin{equation}
 \lim_{\e\to 0}  \R_2(x,\e,\mu) = 
 \frac{3}{x}\int_0^x\dd x'\,x'(x-x')\,S_{R,0}(x',\mu).
 \label{eq:retarded}
\end{equation}
The source-level Laurent test of Sec.~\ref{sec:mechanism} makes the soft limit direct since  Eq.~\eqref{eq:retarded} cannot generate a negative Laurent power when
$S_R$ has none.  We thus reach our first main conclusion:
\begin{equation}
 \big[\nabla^2\R^{(2)}\big]_{\rm LSWN}=0 
 \label{eq:noIR}
\end{equation}
unlike the BIS-II conjecture \eqref{eq:BIS-extrapolation}.
BIS-II requires a nonvanishing value for their observable relic --- a finite white hard-hard response for $\nabla^2 \R^{(2)}$ to be applied to Eq.~\eqref{eq:BIS-R} or an
infrared-divergent $\R^{(2)}$ as $k_L\to0$ --- and it instead vanishes.  The kernel $\R_2$ itself is white, with solutions given in closed form in Eq.~\eqref{eq:R2mono} and \eqref{eq:R2quad}, so the curvature power spectrum contribution generated from hard-hard modes is also white but that makes its IR contributions to curvature modes strongly convergent, not divergent as we shall see.

On the other hand BIS correctly inferred that the kurvature density is white, consistent with the hard-hard shear source in $(\sigma^{(1)})^2$ in Eq.~\eqref{eq:gauss2} or $(\theta^{(1)})^2$ in Eq.~\eqref{eq:kurvature}.
In terms of the  kurvature density normalized to the radiation background $\dK=\Dr/\bar\rho$ these equations become
\begin{equation}
\dK= 
 \dKcomp{density}+\dKcomp{expansion}=\dKcomp{intrinsic}+\dKcomp{shear}.
 \label{eq:allocdK}
\end{equation}
Each of these terms can be evaluated in second-order perturbation theory to form the kurvature kernel.   Explicit substitution of the first-order solutions gives
\begin{equation}
 \begin{aligned}
 \dKcomp{density}&=(E_n)_{AB}/\bar\rho, &
 \dKcomp{expansion}&=-\tfrac19a^2\eta^2(K^2)_{AB},\\
 \dKcomp{intrinsic}&=\tfrac16a^2\eta^2(\threeR)_{AB}, &
 \dKcomp{shear}&=-\tfrac13a^2\eta^2(\sigma^2)_{AB},
 \end{aligned}
 \label{eq:alloc}
\end{equation}
with $\sigma^2\equiv\tfrac12\sigma_{ij}\sigma^{ij}$ the same trace-free-extrinsic-curvature
invariant already used in $K_{ij}=-\tfrac13\theta\gamma_{ij}-\sigma_{ij}$ and
Eq.~\eqref{eq:gaussexact}, evaluated here at $AB$ order rather than at first
order.  Note that $\dKcomp{density}=\delta_2$ and so is the second-order density perturbation in comoving gauge. 
All four terms in the white $\e\to 0$ limit follow:
\begin{align}
 \dKcomp{intrinsic}&=-10\sin^2x ,\label{eq:Ki}\\
 \dKcomp{shear}&=-\tfrac23x^2b(x)^2 ,\label{eq:Ks}\\
 \dKcomp{density}&=-4+\frac{8\sin2x}{x}-\frac{12\sin^2x}{x^2} ,\label{eq:Kd}\\
 \dKcomp{expansion}&=\dKcomp{intrinsic}+\dKcomp{shear}-\dKcomp{density} ,\label{eq:Ke}
\end{align}
which implies
\begin{equation}
 \dK(x)%=\tfrac13\big[K_{\rm intrinsic}(x)+K_{\rm shear}(x)\big]
 =-10\sin^2x-\tfrac23x^2b(x)^2.
 \label{eq:deltaK}
\end{equation}
Note that none of these quantities depend on $\mu$.
We use the Hamiltonian constraint here for the compact displayed form of $\dKcomp{expansion}$, but
have also evaluated it directly from the full ADM $K^2$ in  Appendix~\ref{app:dewitt}.  

Note that despite $k_L^2\R_2$ having identically zero white contribution, the intrinsic curvature does provide a white contribution to $\dK$ but not one that grows  as $a^2 \propto x^2$ in a manner consistent with BIS expectations.  It is simply the hard-hard quadratic terms from the geometric identity Eq.~\eqref{eq:R3geometric}:
$16\R_1 \nabla^2\R_1+6(\partial_i\R_1)^2\to-20\sin^2x/x^2$ for the two antiparallel legs with
$\R_1=\sin x/x$.  With this normalization, 
\begin{equation}
\dKcomp{intrinsic}=\frac{(\threeR)_{AB}}{16\pi G\bar\rho} 
= -20 \frac{\sin^2 x}{x^2} \frac{q^2\eta^2}{6}= -10\sin^2x.
\end{equation}
This expression matches Eq.~\eqref{eq:Ki} exactly and remains bounded at $x\gg 1$.
The white intrinsic $3$-curvature present while the hard acoustic waves are active should not be confused with a propagating long-wavelength curvature mode.  It is carried by the local quadratic hard--hard geometric terms in Eq.~\eqref{eq:R3geometric}, built from $\R_1$ and arising from the same hard acoustic fields that supply the active second-order sources.  The long-wavelength response governed by the master equation is instead $\R_2$, whose Laplacian has identically zero LSWN.  If physical damping removes the hard waves, these local quadratic $3$-curvature terms disappear; any matched $\R_2$ response has $\nabla^2\R_2\propto k_L^2$ and therefore does not yield a BIS-type relic.

The shear term 
$\dKcomp{shear}=-2 \sigma^2_{AB}/16\pi G\bar\rho$ follows similarly from the first-order
 contraction 
 \begin{equation}
(\sigma^2)_{AB} = \frac{2}{3} \frac{q^2 b(x)^2}{a^2}.
\end{equation}
Notice that since $\lim_{x\gg 1} b(x) \rightarrow -\sqrt{3}\sin x$, the shear term supplies a growing $x^2$ contribution to $\dK$, as required by the BIS expectation for LSWN kurvature.  

The density term
$\dKcomp{density}(x)=\delta_2$ is similarly reconstructed from Eq.~\eqref{eq:recon}.  It contains the term 
\begin{multline}
 \eta\dot\R_2=\frac{3}{x}\int_0^x\dd x'\,x'^2S_{R,0}
 =\frac{1}{2x^2}\Big[4x^2\left(1-3\mu^2\right)+24\mu^2-2
 +\left(4x^2\left(3\mu^2-1\right)-24\mu^2+2\right)\cos2x\Big.\\
 \Big.+x\left(3x^2\left(\mu^2-1\right)-24\mu^2+2\right)\sin2x\Big].
 \label{eq:etaR2p}
\end{multline}
The $\mu$ terms cancel with those in $\eta S_M$ in Eq.~\eqref{eq:sourcesclosed} leaving 
\begin{equation}
 \delta_2(x)=-4+\frac{8\sin2x}{x}-\frac{12\sin^2x}{x^2}
 \label{eq:delta2}
\end{equation}
 independently of $\mu$.  Notice that $\lim_{x\gg 1}\dKcomp{density}=-4$ and, like $\dKcomp{intrinsic}$, it stays bounded.
 Notice instead that $\dKcomp{expansion}-\dKcomp{shear}\to-1$ for $x\gg 1$ and is subleading.  As we shall show in Appendix~\ref{app:dewitt}, the common leading $x^2$ growth of the expansion and shear is no coincidence: the nearly antiparallel hard modes require this extrinsic-curvature form.  
 
 In 
summary,
time-averaging over the acoustic phase, so that $\sin^2x\to\tfrac12$ and
$\sin x\cos x\to0$, makes the late-time or UV hierarchy:
\begin{equation}
 \begin{aligned}
 \dKcomp{intrinsic}&\longmapsto-5, &
 \dKcomp{density}&\longmapsto-4,\\
 \dKcomp{shear}&=-x^2+\Order(x^{-2}), &
 \dKcomp{expansion}&=-x^2-1+\Order(x^{-2}),
 \end{aligned}
 \qquad (x\gg1).
 \label{eq:coarseallocation}
\end{equation}
It may be tempting to interpret the $x^2$ term in $\dK$ as an initial condition for a free superhorizon growing mode after the hard waves cross the sound horizon.  But that interpretation would require the quadratic sources to become negligible and the complete second-order system to then match onto  the linear growing-mode relation. Neither occurs in the ideal radiation fluid considered here: the subhorizon acoustic waves retain finite oscillations, so their quadratic sources remain active. The resulting constrained particular solution leaves the comoving density response bounded while its extrinsic expansion and shear contributions grow in $\dK$. The growing scaling of kurvature coinciding with the scaling of the growing linear mode does not make the full second order system  a free linear growing mode.

Instead, the relevant $x^2$-growing white contribution to $\dK$ is an ordinary local effect, not a new
long-range mode: a bounded acoustic beat between the two hard legs
of the extrinsic terms, shear or expansion.   It measures the deformation of the
comoving slicing due to acoustic compression, increasingly rapid relative to the background Hubble expansion.
The intrinsic curvature and comoving density contribute much smaller bounded terms to $\dK$.  Moreover the relativistic constraint system keeps even this small but white \emph{density} term from becoming the $k_L^{-2}$ \emph{curvature} branch that the BIS-II conjecture implies through Eq.~\eqref{eq:BIS-R}, as does the geometric constraint for intrinsic curvature in Eq.~\eqref{eq:R3geometric}.   

Interestingly the LSWN (not IR divergent as with the BIS-II conjecture) comoving-curvature response in Eq.~\eqref{eq:R2mono} and \eqref{eq:R2quad} has a distinct tidal structure:
$\R_2^{\ell=0}\to5/2$, whereas its quadrupole grows as
$\R_2^{\ell=2}\simeq-4\ln(2x)$ for $x\gg 1$.  We discuss this multipole structure in Appendix~\ref{app:multipoles} and show the generation of such a white noise residual does not contradict the exact Langlois--Vernizzi  conservation law in Appendix~\ref{app:lv}. 

\subsection{LSWN power spectra}
\label{sec:P22}

We can now put these considerations together and show why despite LSWN existing for kurvature there is no  IR divergence in curvature.
The direct hard-hard response becomes physically relevant  through the power it
generates at long wavelengths.  We therefore form the hard-hard only $P_{22}$
contribution to the power spectrum.   We then compare it to the same construction implied by the BIS-II conjecture through Eq.~\eqref{eq:BIS-R}.  Note that neither is the complete second-order power spectrum, which involves first--third-order coupling, but this coupling to a pre-existing soft mode is not the focus of LSWN, which arises purely from hard modes. 

We construct the full second-order curvature by integrating over the hard modes.  Recall that our AB coefficients for the hard-hard kernel
$\R_2(x,\epsilon,\mu)  = \mathcal F_2(\bm q,\bm p;\eta)$ when restoring the vector representations of the soft and hard modes 
\begin{equation}
\bm k =\bm q+\bm p.
\end{equation}
Note that for notational simplicity we no longer distinguish the soft mode as $\bm k_L (=\bm k)$ here since we are explicitly calculating LSWN power at $k$ and no further confusion with a generic wavemode should arise.

  The
second-order field is then the integral over all of the allowed hard-hard modes
\begin{equation}
 \R^{(2)}(\bm k,\eta)=\frac12\int\!\frac{\dd^3q}{(2\pi)^3}\,
 \mathcal F_\R(\bm q,\bm k-\bm q;\eta)\,\R_i(\bm q)\R_i(\bm k-\bm q),
 \label{eq:R2conv}
\end{equation}
where the one-half is required because the integral traverses both orderings of the
pair while $\mathcal F_\R$ counts it once.    The second-order power spectrum term for IR modes generated purely from UV modes is the 22 piece:
\begin{equation}
 \big\langle\R^{(2)}(\bm k)\R^{(2)}(\bm k')\big\rangle
 =(2\pi)^3\delta_{\rm D}(\bm k+\bm k')P_{22}(k),
 \qquad
 \Delta^2_{22}(k)=\frac{k^3P_{22}(k)}{2\pi^2} ,
  \label{eq:def22}
\end{equation}
where
\begin{equation}
 P_{22}(k,\eta)=\frac12\int\!\frac{\dd^3q}{(2\pi)^3}\,
 \big|\mathcal F_\R(\bm q,\bm k-\bm q;\eta)\big|^2
 P_{\R_i}(q)\,P_{\R_i}(|\bm k-\bm q|).
 \label{eq:P22}
\end{equation}
The full second-order power spectrum would contain
 $\langle\R^{(1)}\R^{(3)}\rangle$ but those require the existence of a primordial IR mode and are not generated purely from hard-hard modes.  Notice that if the kernel contains a white $\e^0$ form then $P_{22}(k,\eta)$ has a constant or white noise form for $\e=k/q \ll 1$ for any initial power spectrum with support for such $q$.
 
\begin{figure}[!htbp]
\centering
\includegraphics[width=0.62\textwidth]{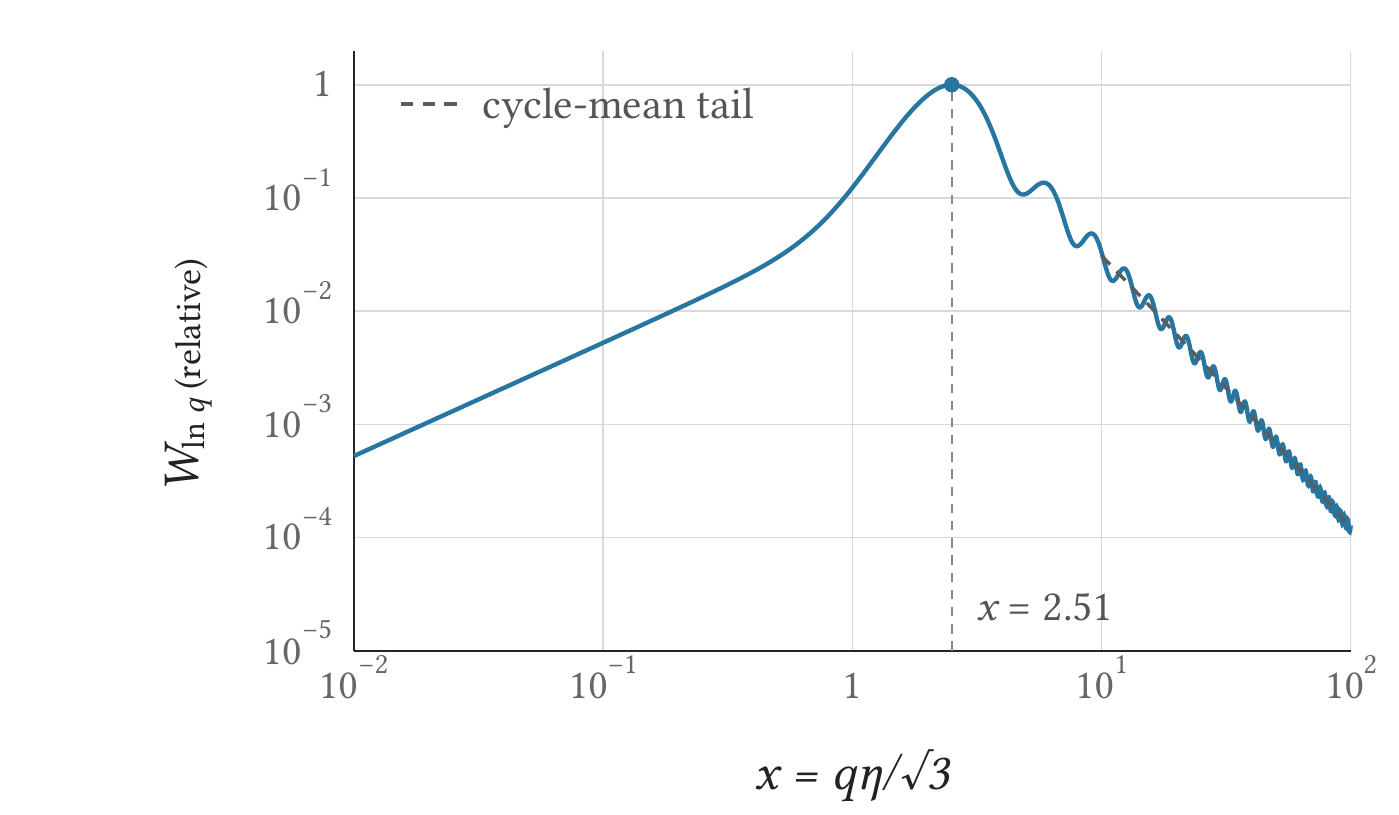}
\caption{Relative weight with which hard modes contribute to the hard-hard LSWN curvature power spectrum, $\mathrm{d}\Delta_{22}^2/\mathrm{d}\ln q \propto W_{\ln q}$, at a fixed evaluation epoch $\eta$.   The weight is both strongly IR and UV convergent and peaks for modes that just cross the sound horizon $x =2.51$.  Unlike the BIS-II conjecture, there is no UV cutoff sensitivity to $\Delta_{22}^2$.  The cycle-mean tail (dashed line) is an analytic result from Eq.~\eqref{eq:Wlnqtail}.}
\label{fig:loghw}
\end{figure}

 To make these considerations concrete, let us specialize to  a scale-invariant initial spectrum, $P_{\R_i}=2\pi^2A_s/q^3$ or $n_s=1$.  Then
 \begin{equation}
 \Delta^2_{22}(k,\eta)=\tfrac12A_s^2\,k^3\!\int\!\dd q\;q^{-4}
 \big\langle|\mathcal F_\R|^2\big\rangle_\mu ,
 \label{eq:P22weight}
\end{equation}
so each hard mode enters with weight $q^{-3}|\mathcal F_\R|^2$ per logarithmic
interval.  Given the monopole--quadrupole split
$\mathcal F_\R=\R_2^{\ell=0}(x)+\R_2^{\ell=2}(x)P_2(\mu)$ of
Eqs.~\eqref{eq:R2mono}--\eqref{eq:R2quad} we average over angles: $\langle P_2\rangle_\mu=0$ and
$\langle P_2^2\rangle_\mu=1/5$.  This  gives
\begin{equation}
\langle|\mathcal F_\R|^2\rangle_\mu=\big(\R_2^{\ell=0}\big)^2+\big(\R_2^{\ell=2}\big)^2/5,
\end{equation}
so Eq.~\eqref{eq:P22weight} becomes
\begin{equation}
 \Delta^2_{22}(k,\eta)=A_s^2(k\eta)^3 I_W,
\end{equation}
where the integral
\begin{equation}
I_W\equiv
  \int\dd(\ln x) \,W_{\ln q}(x),
 \qquad
 W_{\ln q}(x)\equiv\frac{\big(\R_2^{\ell=0}\big)^2+\big(\R_2^{\ell=2}\big)^2/5}{6\sqrt3\,x^3} ,
 \label{eq:Wlnq}
\end{equation}
and recall  $x=q\eta/\sqrt3$, where we now consider it at fixed $\eta$ and varying $q$.  Note that if read as a function of $\eta$, this form requires $k\eta\ll 1$ for superhorizon LSWN.
Here $W_{\ln q}$ is the contribution of each logarithmic hard-wavenumber interval.  
We plot the relative weight in Fig.~\ref{fig:loghw} using 
the full 
kernels not their $x\gg1$ limits.  
The weight is regular as $x\to0$ and peaks at $x\simeq2.51$.  At high $x$ the cycle-averaged tail falls as $x^{-3}\log^2(2x)$ (see Appendix~\ref{app:multipoles}).   Together these imply that
 the modes that contribute most strongly to $\Delta_{22}^2$ for a scale-invariant initial power spectrum are those that cross the sound horizon at the epoch $\eta$ at which the power spectrum is evaluated.

\begin{figure}[!htbp]
\centering
\includegraphics[width=0.72\textwidth]{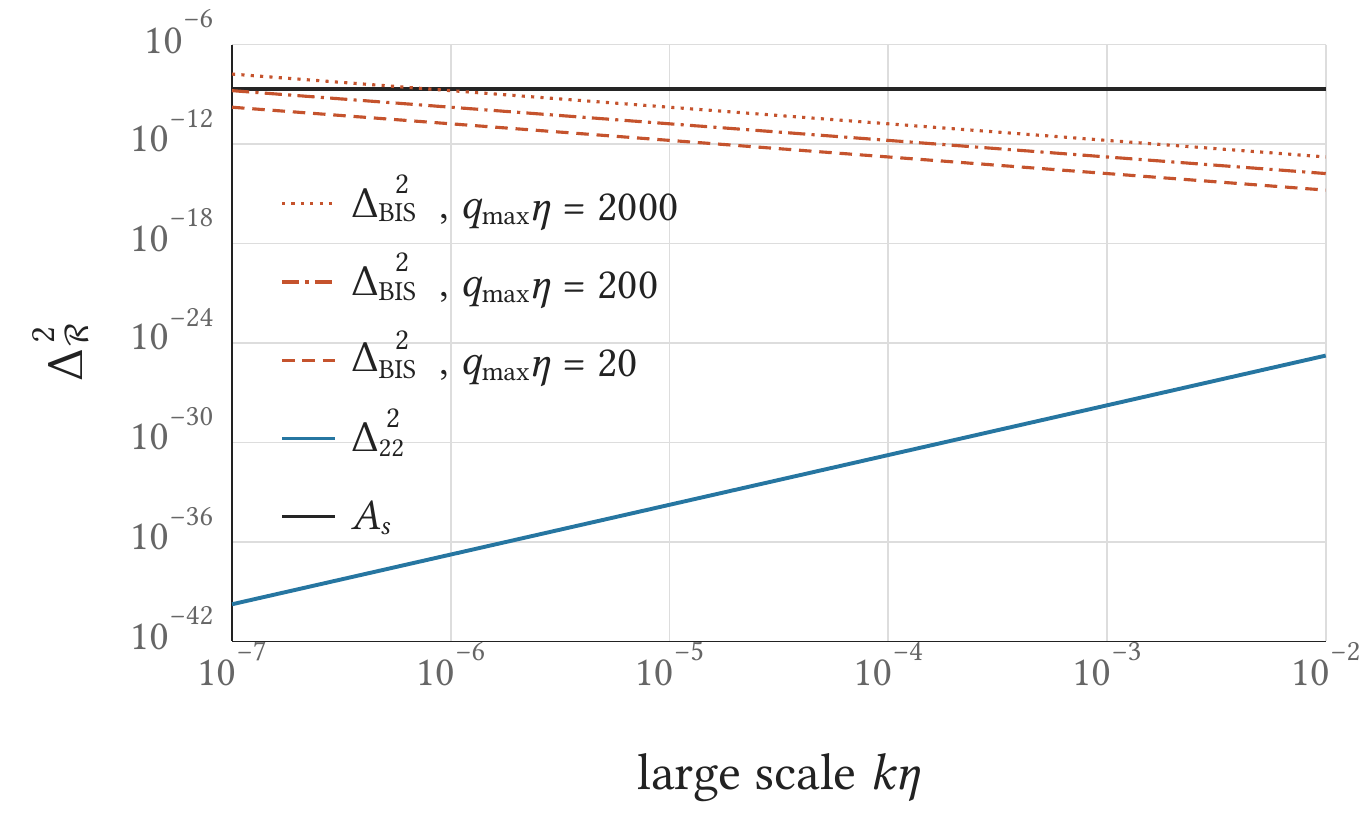}
\caption{LSWN curvature power spectra for the hard-hard contributions at second order $\Delta_{22}^2$ compared with that inferred by the BIS-II conjecture $\Delta_{\rm BIS}^2$ which relates kurvature density to curvature through a Poisson-like equation \eqref{eq:BIS-R}, as well as the initial scale-invariant value of $A_s=2.1\times 10^{-9}$.   Our $\Delta_{22}^2$ shows no UV cutoff sensitivity from integrating hard modes to $q_{\rm max}$ consistent with the weight shown in Fig.~\ref{fig:loghw}.   It is strongly suppressed as $(k\eta)^3$ on superhorizon scales.   Conversely $\Delta_{\rm BIS}^2 \propto q_{\rm max}/k$ and shows cutoff-sensitive IR divergence.}
\label{fig:cutoff-comparison}
\end{figure}

 A numerical integration of the weight gives
$I_W \approx 0.03918$,
and with this value, we  plot the final result for LSWN in curvature in Fig.~\ref{fig:cutoff-comparison} assuming $A_s=2.1\times 10^{-9}$.  Notice that LSWN curvature is strongly IR convergent, with $\Delta_{22}^2$  rising to the horizon scale as $(k\eta)^3$.   Only near the horizon scale itself do we even approach the expected small-22 $\Order(A_s)$ relative correction to the power spectrum.

Contrast this with the BIS-II inferred response in Eq.~\eqref{eq:BIS-R}, given our calculation of kurvature LSWN:
\begin{equation}
\R_{\rm BIS}\big|_{AB}=\frac{4\pi Ga^2\bar\rho(\eta)}{k^2}\,\dK(x)
 =\frac{3}{2} \frac{\dK(x)}{(k\eta)^2} .
\end{equation}
With Eq.~\eqref{eq:deltaK}, the kernel then becomes
\begin{equation}
 \mathcal F_{\rm BIS}(\bm q,\bm p;\eta)\equiv \frac{q^2}{k^2 }
\left[-5\frac{\sin^2x}{x^2} -\tfrac13b(x)^2\right].
 \label{eq:BISkernel}
\end{equation}
Notice that it is only the second, or shear term, in brackets that can masquerade as a constant mode at late times.  The first or intrinsic curvature term, even by the BIS accounting, would decay.
This kernel gives the BIS power spectrum for scale-invariant initial spectra
\begin{equation}
 \Delta^2_{\rm BIS}(k,\eta)=\tfrac12A_s^2\!\int\!\dd \ln q \left(\frac{k}{q}\right)^3
 \big\langle|\mathcal F_{\rm BIS}|^2\big\rangle_\mu .
 \label{eq:BISweight}
\end{equation}
The square of the kernel provides $q^4/k^4$, so the integral up to  $q_{\rm max}$ is UV sensitive and its impact is IR divergent:
\begin{equation}
 \Delta^2_{\rm BIS}(k,\eta)\simeq\frac{3}{16}
\frac{q_{\max}}{k}A_s^2 . 
 \label{eq:BISasymptotic}
\end{equation}
In Fig.~\ref{fig:cutoff-comparison}, we plot this BIS prediction for several values of $q_{\max}\eta$.   Notice that for sufficiently large values, the BIS spectrum crosses the primordial spectrum on superhorizon scales.   This is the basis of the observational bound on $q_{\max}$ in BIS-II.   No such IR-divergent curvature power spectrum actually occurs with an explicit second-order calculation of $\Delta_{22}^2$, where the dominant contributions are from horizon-scale modes and provide only the expected $\Order(A_s)$ fractional correction at the edge of the superhorizon regime.   The direct observable in the CMB requires a Boltzmann calculation of the multicomponent cosmological system.  Nevertheless,  this calculation already shows the absence of an IR enhancement of the Sachs-Wolfe effect, giving instead a strong $(k\eta)^3$ suppression of $\Delta_{22}^2$ and hence its impact on low order multipoles.

\section{Discussion}
\label{sec:disc}

We have shown by explicit calculation of the second-order evolution of acoustic waves during radiation domination: where LSWN lives and where it does not, and what contributes to kurvature and what does not.
Beyond linear theory, kurvature contains both intrinsic 3-curvature and extrinsic curvature through the quadratic composite of first-order shear.   The growing contribution to the fractional kurvature density $\dK$, which  mimics that of the matter density $\delta$ above the horizon and provides its LSWN, comes exclusively from the shear of nearly antiparallel acoustic beat modes.

The much smaller intrinsic 3-curvature contribution is also constructed from beat coupling of first-order curvature modes and has identically zero LSWN in the Laplacian of the curvature mode, invalidating the BIS-II conjecture of an IR divergence in observable curvature modes. 
Geometrically, the curvature mode $\R$ is not the potential for 3-curvature fluctuations beyond linear theory.
Through the Hamiltonian constraint, these results can be recast exactly in terms of the comoving density fluctuation and the quadratic composite of first-order expansion.  Here the growing contribution to $\dK$ lives exclusively in the expansion term, again an extrinsic-curvature contribution, whereas the second-order density reaches a constant soon after horizon crossing of the hard mode.  This relationship between the shear and expansion views of the source of kurvature also provides geometric insight: they represent the nearly plane-parallel distortions of the comoving slice due to the acoustic compression induced by nearly antiparallel beat-coupled modes. 

As a consequence, the hard-hard only contributions to the second-order curvature power spectrum are white, not IR divergent and strongly convergent, and never exceed the first-order primordial power spectrum.  Their weight mainly comes from hard modes that are on the horizon scale at the evaluation epoch, exactly as one might expect from relativistic terms in second-order perturbation theory.   The absolute correction reaches the naive second-order estimate of $A_s^2$ only when the soft mode approaches the horizon scale as well, and is never important at superhorizon or LSWN scales.

The ideal radiation fluid used here retains undamped acoustic oscillations, so its quadratic hard-hard sources do not become negligible merely when the hard modes cross the sound horizon.  The direct response is therefore a continuously sourced particular solution, rather than automatically a free growing or decaying mode.  If a physical damping process or transition instead suppresses the sources, the subsequent long-wavelength response can be matched onto free solutions.  In the LSWN limit, writing the source envelope as $W(x)$, Eq.~\eqref{eq:retarded} gives
\begin{equation}
 \R_2(x)=C_++\frac{C_-}{x},\qquad
 C_+=3\int_0^\infty\dd x\,xS_{R,0}(x)W(x),\qquad
 C_-=-3\int_0^\infty\dd x\,x^2S_{R,0}(x)W(x),
 \label{eq:matched-relic}
\end{equation}
after the sources have switched off.  Such matching can leave a finite constant curvature component as well as a decaying one.  It does not restore the BIS-II result: the hard-hard source is regular as $k_L\to 0$, so any matched remnant is a finite white response rather than a $k_L^{-2}$ curvature potential.

 While this example provides an instructive counterexample to the BIS-II conjecture and directly addresses the claims of an observational LSWN effect in the CMB, several open questions remain.  First we have only considered the impact of hard-hard modes alone as appropriate for LSWN and not the full second-order power spectrum, which must include the first--third-order term or $\Delta^2_{13}$.  Second, we follow BIS in assuming an irrotational fluid and so do not address the effect of the rotation term $\omega^2$ on kurvature.   
 
Finally, our finding of a small white-noise $k^0$ term in the second-order comoving density $\delta_2$ and curvature fluctuations $\R_2$ might seem to contradict naive Newtonian gravity expectations of only $k^4$ power due to energy and momentum conservation \cite{Zeldovich:1969sb,Peebles:1980yev}.  The curvature fluctuation even has a small logarithmically growing tidal term from UV modes.  However, the net contribution to the power spectrum comes mainly from hard modes that are on the horizon at the time, and it is well known that relativistic slicing and threading choices can correct the naive expectation in a gauge-dependent way \cite{Bartolo:2005xa}.  Furthermore, we have found that in comoving gauge, subhorizon hard modes do contribute white-noise terms to the local expansion, reminiscent of backreaction questions but in the domain of perturbation theory rather than nonlinear structure.  
Moreover the second-order shift potential does contain an inverse Laplacian. Although the associated ADM shift is infrared integrable, its power is less infrared-suppressed than that of the linear shift. In an individual horizon-sized realization, this long-wavelength shift could provide directionally biased local threading even though the ensemble is statistically isotropic. 
%Whether such fluctuations survive a physically defined coarse graining as an anisotropy of the local background is a separate question.

For all of these issues that relate to the validity of an approximate FLRW expansion in a given horizon-sized realization, one must be careful to define a coarse-grained description rather than one that follows the nonlinearly compressing small-scale acoustic fluctuations themselves.  Determining whether these local fluctuations alter the coarse-grained FLRW expansion requires a separate averaging/backreaction analysis; by themselves, they do not establish such an effect (see e.g.~\cite{Green:2010qy} and references therein).
 We leave these questions for future study.

\section*{Acknowledgments}

We thank Keisuke Inomata, Austin Joyce and Hayden Lee for useful conversations and Albert Stebbins for helpful discussions and comments on a draft version. WH was supported 
 by U.S.\ Dept.\ of Energy contract DE-SC0009924 and the Simons Foundation. 
The author used OpenAI Codex (GPT-5; accessed July 2026) and Anthropic Claude
(including models through Claude Opus 5) as interactive assistants for literature
organization, manuscript and \LaTeX\ preparation, and coding and figure-workflow
support.  The author directed, checked, and revised all outputs used here.  All
scientific judgments, derivations, calculations, interpretation, and responsibility
for the manuscript remain solely with the author.

\appendix
\section{Extrinsic curvature of the hard-hard response}
\label{app:dewitt}

In Sec.~\ref{sec:Adivergence} we identified the common growth of the $\dK$ expansion and shear contributions, $\dKcomp{expansion}$ and $\dKcomp{shear}$ in Eq.~\eqref{eq:coarseallocation},  as
the kinematics of longitudinal acoustic compression and rarefaction.  This appendix
first gives that plane-wave argument explicitly, then its exact, gauge-covariant
DeWitt form, and the direct evaluation of $\dKcomp{expansion}$ as opposed to inferring it from the Hamiltonian constraint.  It also identifies the bounded difference of the expansion and shear terms  as the
background--second-order contribution to the gravitational kinetic term.

\subsection{Longitudinal acoustic compression and rarefaction}
\label{app:plane}

A plane acoustic wave is a longitudinal, alternating compression and rarefaction
of the fluid.  For a wave along $\hat z$ the velocity is $v(z,\eta)\hat z$, so
the deformation-rate tensor has a single non-vanishing eigendirection,
\begin{equation}
 D_{ij}=\lambda\,\hat z_i\hat z_j .
\end{equation}
The transverse directions are untouched.  This one-dimensional longitudinal
deformation is simultaneously an expansion or compression and a shear, in a fixed
ratio:
\begin{equation}
 \theta=\lambda,\qquad
 \sigma_{ij}=\lambda\,\mathrm{diag}\!\left(-\tfrac13,-\tfrac13,\tfrac23\right),
 \qquad \sigma_{ij}\sigma^{ij}=\tfrac23\theta^2 ,
\end{equation}
and hence
\begin{equation}
 K_{ij}K^{ij}=\tfrac13\theta^2+\sigma_{ij}\sigma^{ij}=\theta^2=K^2 .
\end{equation}
The Hamiltonian constraint~\eqref{eq:ham} contains only
$K^2-K_{ij}K^{ij}$, so a single plane wave contributes nothing to it, however
large its compression.

A soft $k_L \ll q$ from two hard modes requires $\bm p=\bm k_L-\bm q$ and therefore,
in the squeezed limit, antiparallel waves:
$\hat {\bm q}\cdot\hat {\bm p}=-1+\Order(\e^2)$.  They deform the fluid along the same axis, so their
combined longitudinal deformation is again one-dimensional.  This is the physical reason the
$x^2$ growth cancels between the expansion and shear entries of the Hamiltonian constraint.

\subsection{DeWitt form of the hard-leg cancellation}
\label{app:dewitt-null}

The preceding discussion phrases the cancellation in the fluid picture while our extrinsic curvature is a property of the embedded spatial slice.  In the
ADM foliation used in Sec.~\ref{sec:ham}, the spatial projector is
$h_{\mu\nu}=g_{\mu\nu}+n_\mu n_\nu$.  The induced metric on a constant-time
hypersurface is its spatial pullback,
$\gamma_{ij}=h_{\mu\nu}e^\mu_i e^\nu_j$ for tangent vectors $e^\mu_i$ to that
slice, and $K_{ij}$ is minus one half of its normal rate of change in the ADM
convention of Sec.~\ref{sec:ham}.
The longitudinal compression and accompanying shear are therefore components of
the same $K_{ij}$.  Their invariant combination in the Hamiltonian constraint is
the gravitational kinetic term,
\begin{equation}
 \mathcal G_{\rm DW}^{ijkl}K_{ij}K_{kl} =  K_{ij}K^{ij}-K^2 ,
 \label{eq:DeWitt}
\end{equation}
where the 
\begin{equation}
 \mathcal G_{\rm DW}^{ijkl}=\tfrac12\left(\gamma^{ik}\gamma^{jl}+\gamma^{il}\gamma^{jk}-2\gamma^{ij}\gamma^{kl}\right),
\end{equation}
is the DeWitt supermetric.
Its one negative trace direction and five positive trace-free directions encode
the competition between expansion and shear.  At the order needed below, indices
are raised and lowered with the background $\delta_{ij}$.  The corresponding
DeWitt bilinear form is
\begin{equation}
 \mathcal G_{\rm DW}(v,w)\equiv\mathcal G_{\rm DW}^{ijkl}v_{ij}w_{kl}
 =v_{ij}w^{ij}-(\operatorname{tr}v)(\operatorname{tr}w) ,
 \qquad\text{so that}\qquad K_{ij}K^{ij}-K^2=\mathcal G_{\rm DW}(K,K) .
 \label{eq:Gbracket}
\end{equation}

In the comoving gauge of Sec.~\ref{sec:radiation}, the first-order extrinsic
curvature perturbation on a single hard leg is
\begin{equation}
 \delta K^i{}_j=-\frac1a\Big[\big(\dot\R-\Hc\lapse\big)\,\delta^i_j+k^ik_jB\Big] .
 \label{eq:dKfull}
\end{equation}
The isotropic piece is exactly the linear momentum constraint,
$\dot\R-\Hc\lapse=0$ (Eq.~\eqref{eq:linM}), so it vanishes identically rather than
only at leading order, leaving
\begin{equation}
 \delta K^i{}_j=-\frac{k^ik_jB}{a}=L\,\hat k^i\hat k_j ,
 \qquad L\equiv-\frac{k^2B}{a} ,
 \label{eq:dKrank1}
\end{equation}
a rank-one tensor.  For a rank-one outer
product of a unit vector, $\operatorname{tr}v=(v_{ij}v^{ij})^{1/2}$, so the
identities collapse trivially: $v_{ij}=L\hat k_i\hat k_j$ gives $\operatorname{tr}v=L$
and $v_{ij}v^{ij}=L^2$, so
\begin{equation}
 \mathcal G_{\rm DW}(\delta K_q,\delta K_q)=L_q^2-L_q^2=0
\end{equation}
and similarly for the $p$ leg.  Two null directions need not have a vanishing cross term,
but strict-soft kinematics forces the two hard-leg deformations onto the same
generator: $\bm p=\bm k_L-\bm q$ gives
$\hat {\bm q}\cdot\hat {\bm p}=-1+\Order(\e^2)$, hence $\hat {\bm p}_i\hat {\bm p}_j=\hat {\bm q}_i\hat {\bm q}_j+\Order(\e^2)$.
The two rank-one tensors are therefore proportional up to the soft misalignment,
and
\begin{equation}
 \mathcal G_{\rm DW}(\delta K_q,\delta K_p)=\Order\!\left(\e^2|\delta K_q|\,|\delta K_p|\right) .
 \label{eq:Gcross}
\end{equation}
Thus the leading hard-hard contribution cancels in the gravitational kinetic
term.  This null property is specific to the comoving slicing: in synchronous
gauge the corresponding $\delta K^i{}_j$ has an isotropic component and is not
null.

\subsection{Direct computation of $\dKcomp{expansion}$}

The direct computation of the expansion term in the kurvature is operationally more cumbersome since unlike the shear, it enters with a background contribution and requires a higher order expansion.   In the main text, we took the short cut of using the Hamiltonian constraint to infer it from the shear term.  To avoid the impression of circularity in the discussion of shear and expansion, we carry through the direct computation here.

The background extrinsic curvature is pure trace, $\bar K^i{}_j\propto\delta^i_j$,
the opposite extreme from a null direction:
$\mathcal G_{\rm DW}(\bar K,\bar K)/|\bar K|^2=-2$ exactly.  It sits on the timelike (volume) axis
of the DeWitt cone in Eq.~\eqref{eq:DeWitt}, orthogonal in kind to the null
leg deformations above.  The $AB$ coefficient of the kinetic term therefore
splits into a leg--leg piece and a background--second-order piece,
\begin{equation}
 \big(K_{ij}K^{ij}-K^2\big)_{AB}={2\,\mathcal G_{\rm DW}(\delta K_q,\delta K_p)}
 \;+\;{2\,\mathcal G_{\rm DW}(\bar K,K|_{AB})}
 =  \frac{4\Hc}{a}K\big|_{AB} + \Order(\e^2),
 \label{eq:kinsplit}
\end{equation}
so essentially the entire bounded difference $K_{ij}K^{ij}-K^2$ at $AB$ order is
carried by $K|_{AB}$, the trace of the second-order extrinsic curvature.

From $K=-\tfrac1N\big[\dot{(\ln\sqrt\gamma)}-D_i\beta^i\big]$, where $D_i$ is
the covariant derivative compatible with the induced metric and
$D_i\beta^i\equiv\gamma^{-1/2}\partial_i(\gamma^{1/2}\beta^i)$, with
$\beta^i\equiv\gamma^{ij}\beta_j$, $\gamma_{ij}=a^2(1+2\R)\delta_{ij}$,
$\beta_i=a^2\partial_iB$, and
$N^2=a^2(1+2\lapse)+\gamma^{ij}\beta_i\beta_j$, the $AB$ coefficient of $-aK$ is,
to second order 
\begin{equation}
 \begin{aligned}
 -a\,K\big|_{AB}=\;&3\dot\R_2+k_L^2B_2-3\Hc\lapse_2\\
 &-6\big(\R_q\dot\R_p+\R_p\dot\R_q\big)
 -2\big(p^2\R_qB_p+q^2\R_pB_q\big)
 +(\bm q\cdot\bm p)\big(\R_qB_p+\R_pB_q\big)\\
 &-\lapse_q\big(3\dot\R_p+p^2B_p\big)-\lapse_p\big(3\dot\R_q+q^2B_q\big)
 +3\Hc\big[(\bm q\cdot\bm p)B_qB_p+3\lapse_q\lapse_p\big] ,
 \end{aligned}
 \label{eq:KAB}
\end{equation}
with the first line linear in the second-order fields and the rest quadratic in
the first-order legs; $k_L^2B_2$ survives the soft limit because $B_2\sim k_L^{-2}$.
Inserting the reconstructed $\R_2,\lapse_2,B_2$ of Sec.~\ref{sec:mechanism} into
Eq.~\eqref{eq:KAB} and forming $(K^2)_{AB}$ \emph{directly} gives the same $\dKcomp{expansion}$ as the Hamiltonian constraint.

\section{Comoving curvature and density of the hard-hard response}
\label{app:multipoles}

Both the comoving curvature and density modes acquire a white response to hard-hard modes.  Unlike the shear and expansion terms, they carry no growing contribution to $\dK$ that would indicate a BIS-relation-type effect; their behavior is nevertheless worth examining separately, especially in contrast to linear-theory expectations from conservation laws.

For the curvature mode, Equation~\eqref{eq:retarded} can be broken up into  monopole and quadrupole terms,
\begin{equation}
 \R_2(x,\mu)=\R_2^{\ell=0}(x)+\R_2^{\ell=2}(x)\,P_2(\mu),
 \end{equation}
 where
 $P_2(\mu)=\tfrac12(3\mu^2-1)$,
\begin{align}
 \R_2^{\ell=0}={}&2+\cos^2x-\frac{3\sin^2x}{x^2} \nonumber\\
 & \;\longrightarrow\;\tfrac52 ,
 \label{eq:R2mono}\\[3pt]
 \R_2^{\ell=2}={}&-4\ln(2x)+\sin^2x+4\,\mathrm{Ci}(2x)-4\gamma_{\rm E}+8
 -\frac{8\sin^2x}{x^2}\nonumber\\
& \;\longrightarrow\;
 -4\ln(2x)+\tfrac{17}{2}-4\gamma_{\rm E},
 \label{eq:R2quad}
\end{align}
and the $\rightarrow$ represents taking the cycle average of the $x\gg1$ limit.
The angular structure is particularly transparent in Eq.~\eqref{eq:etaR2p}: its dependence is at most linear in \(\mu^2\), and therefore decomposes entirely into monopole and quadrupole terms.
The monopole is simple  and bounded.  \emph{All} of the secular growth sits in the
quadrupole.  
We conclude that a pair of counter-propagating hard acoustic waves produces a bounded isotropic
curvature response and a logarithmically growing \emph{tidal} one.  This logarithm is a feature of the continuously driven ideal-radiation solution: if a physical process suppresses the hard-hard source, the subsequent strict-soft response takes the free constant-plus-$1/x$ decaying form of Eq.~\eqref{eq:matched-relic}, so the logarithmic growth ceases and fixes a finite constant component together with a decaying one.

\begin{figure}[!htbp]
\centering
\includegraphics[width=0.68\textwidth]{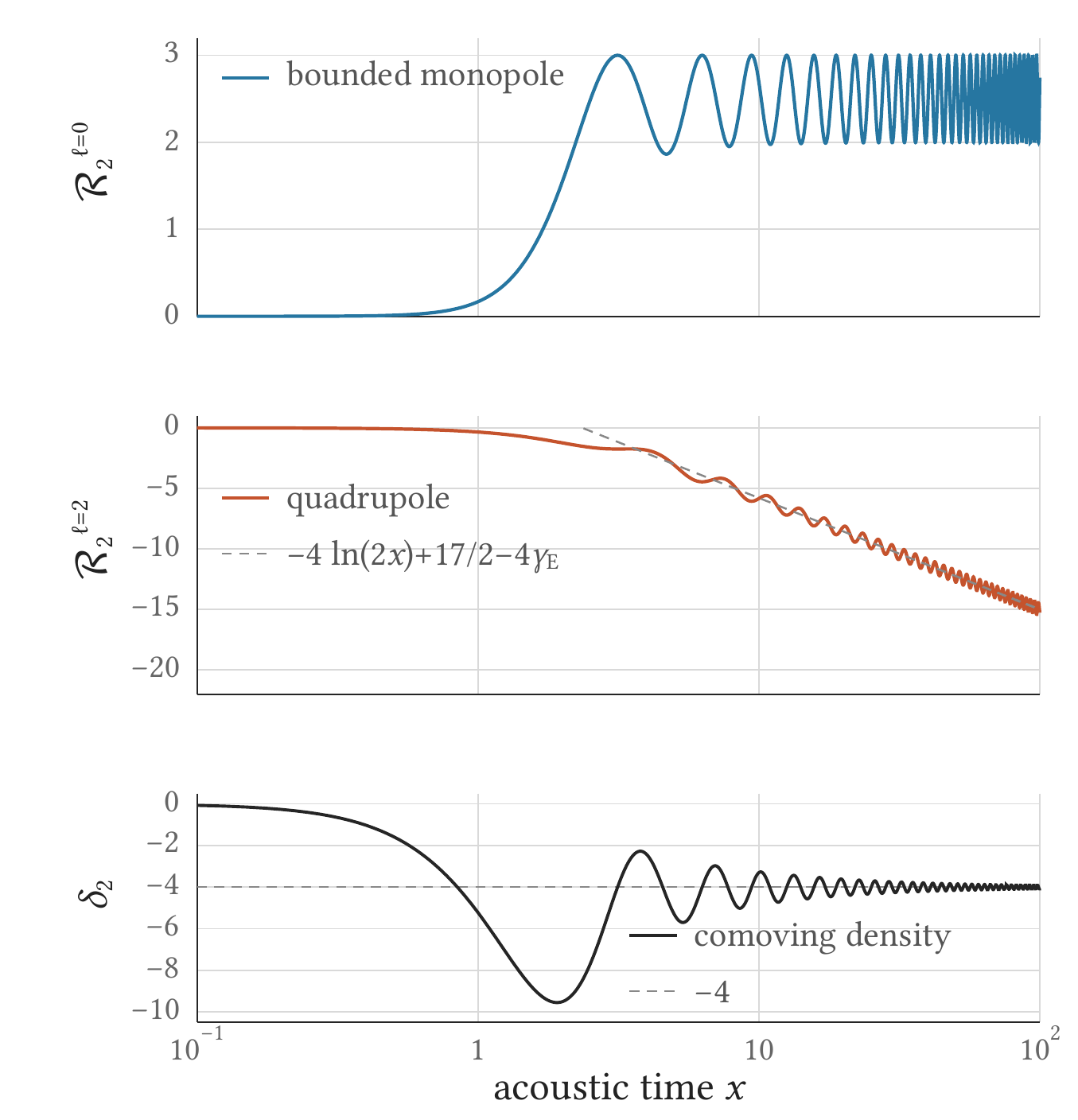}
\caption{Evolution of the curvature and density hard-hard mode response.    Top: the monopole $\R_2^{\ell=0}$ (Eq.~\eqref{eq:R2mono}),
bounded and oscillating about its late-time cycle mean $\tfrac52$.  Middle: the
quadrupole $\R_2^{\ell=2}$ (Eq.~\eqref{eq:R2quad}), with the dashed guide line
$-4\ln(2x)+\tfrac{17}{2}-4\gamma_{\rm E}$ showing the late-time cycle mean.  Bottom: the
comoving density $\delta_2$ (Eq.~\eqref{eq:delta2}), with oscillations that decay toward its asymptotic value $-4$.}
\label{fig:multipoles}
\end{figure}

This decomposition also allows us to calculate the large $x$ limit of $W_{\ln q}$ in Eq.~\eqref{eq:Wlnq}, the weight of the hard modes contributing to the 
second-order $\Delta_{22}^2$ spectrum.
  At large $x$, dropping the transient cosine integral
$\mathrm{Ci}(2x)$ and $1/x^2$ pieces of
Eqs.~\eqref{eq:R2mono}--\eqref{eq:R2quad} leaves
$\R_2^{\ell=0}\to2+\cos^2x$ and $\R_2^{\ell=2}\to\Xi(x)+\sin^2x$ with
$\Xi(x)\equiv-4\ln(2x)+8-4\gamma_{\rm E}$, where $\gamma_{\rm E}$ is Euler's
constant; cycle-averaging their squares
($\langle\cos^2\rangle=\langle\sin^2\rangle=\tfrac12$,
$\langle\cos^4\rangle=\langle\sin^4\rangle=\tfrac38$) gives the closed-form
envelope plotted as the dashed line in Figure~\ref{fig:loghw},
\begin{equation}
 W_{\ln q}^{\rm cycle}(x)=\frac{1}{6\sqrt3\,x^3}
 \left[\frac{51}{8}+\frac15\Big(\Xi(x)^2+\Xi(x)+\frac38\Big)\right] ,
 \label{eq:Wlnqtail}
\end{equation}
which falls as $x^{-3}\log^2(2x)$ up to bounded terms.  

The comoving density follows the same pattern: by Eq.~\eqref{eq:delta2}
its quadrupole vanishes identically and it is pure monopole and bounded. Here $\delta_2=4S_E-4\eta\dot\R_2+4\sqrt3xS_M$, and we see that the
$\mu^2$ (quadrupole) part of $\eta\dot\R_2$ is identical to that of $\sqrt3xS_M$ from Eq.~\eqref{eq:SM}, so it
cancels in the sum, while $S_E$ carries no $\mu$-dependence at all --- leaving the
exactly $\mu$-independent, pure-monopole $\delta_2$ of Eq.~\eqref{eq:delta2}.  The piece of $\R_2$ that grows logarithmically is exactly the piece that the density's own
defining combination is built to eliminate.  Thus no logarithmic terms enter $\dKcomp{density}$ or $\dKcomp{intrinsic}$, the latter because $k_L^2\R_2$ contains no white noise term.  

\section{Synchronous-gauge construction of the hard-hard response}
\label{app:synch}

Comoving gauge results for the hard-hard response can equivalently be constructed by evolving the perturbations
in synchronous gauge using previously established work and transforming the result to the scalar-comoving,
isotropically threaded gauge.  This appendix develops that representation as a cross-check of results in the main paper.   The synchronous gauge solution follows the same general procedure as the comoving gauge derivation in Sec.~\ref{sec:radiation} for its own metric variables but then requires a pullback to the comoving slicing and threading.
  The
pullback is essential: synchronous and comoving curvature variables are not
themselves the same object.  Once expressed in the comoving gauge, the curvature
response is the same $\R_2$ as in Sec.~\ref{sec:radiation} and contains no IR divergent response.

\subsection{Synchronous variables and evolution}
\label{app:synch-source}

For a hard leg of comoving wavevector $\bm b$ ($b\in\{q,p\}$) with unit primordial
curvature, write $x_b\equiv|\bm b|\,\eta/\sqrt3$, so that $x_b=x$ for the
$q$ leg.  We follow the synchronous-gauge notation and results of Inomata, Lee, and Hu (ILH)
\cite{Inomata:2023faq} rather than reproducing them here: $\Psi_b$ and $\hat E_b$ are respectively the isotropic and
scalar trace-free parts of the spatial metric, and $\hat v_b$ is the velocity
potential.  We denote ILH's soft external wavevector by $k_L$, reserving $q$ and
$p$ for the hard legs.  Second-order sources are formed from the first-order solutions,
\begin{equation}
 \Sigma_b\equiv\Psi_b+\frac13\hat E_b=2\,\frac{\cos x_b-1}{x_b^2} ,
 \label{eq:Sigmab}
\end{equation}
together with the velocity-potential transfer
\begin{equation}
 \hat v_b=\frac{\sqrt3}{|\bm b|\,x_b}\Big(2\cos x_b+x_b\sin x_b-2\Big) .
 \label{eq:vb}
\end{equation}
The first-order spatial metric perturbation on each leg is $C^b_{ij}=-\Psi_b\delta_{ij}
+\hat E_b\big(\hat b_i\hat b_j-\tfrac13\delta_{ij}\big)$, and spatial derivatives
act algebraically, $\partial_iQ_b\to i b_iQ_b$.

ILH then constructs the second order sources with the same iterative approach as in Sec.~\ref{sec:radiation}, with the exception that all 4 Einstein equations are used  (their Eqs.~B.24--B.27) instead of the Einstein-Euler system.
The published ILH trace source omits the scalar velocity-squared term $v_iv_i$,
the isotropic trace of the quadratic velocity tensor in the spatial Einstein equation.
This typographical omission is corrected in an arXiv v3 update of ILH.  Correcting it in their Eq.~B.27,
the additional term in the source convention used here is
\begin{equation}
 \Delta S_{kk}=6\Hc^2(1+w)v_iv_i\;\xrightarrow{\ w=1/3\ }\;
 \Delta S_{kk}=8\Hc^2v_iv_i ,
 \label{eq:Skk-fix}
\end{equation}
which is included in $S_{kk}$ below.  We write the scalar metric variable $\hat E$
itself, avoiding a convention-dependent Laplacian redefinition.

Combining the trace and density equations eliminates $\delta$ and gives the
isotropic evolution equation, their Eq.~B.29,
\begin{equation}
 \ddot\Psi+(2+3w)\Hc\dot\Psi-\tfrac19(1+3w)\nabla^2(3\Psi+\hat E)=\tfrac16\left(S_{kk}+3wS_{00}\right).
% \;\xrightarrow{\ 3w=1\ }\;
 %\tfrac16\left(S_{kk}+S_{00}\right) .
 \label{eq:B29}
\end{equation}
Specializing Eq.~\eqref{eq:B29} to radiation and including the trace-free equation gives
\begin{align}
 \ddot\Psi_2+3\Hc\dot\Psi_2+\tfrac29k_L^2\left(3\Psi_2+\hat E_2\right)&=\tfrac16\left(S_{kk}+S_{00}\right)\Big|_{AB},
 \label{eq:Psi2evol}\\
 \ddot{\hat E}_2+2\Hc\dot{\hat E}_2-k_L^2\left(\Psi_2+\tfrac13\hat E_2\right)&=\hat N_{ij}S_{ij}\big|_{AB},
 \label{eq:E2evol}
\end{align}
with $\hat N_{ij}$ the transverse-traceless projector and $S_{kk}$ carrying the
term in Eq.~\eqref{eq:Skk-fix}.  
%The regular solution is selected at early times;
%no independent second-order homogeneous mode is added.

Once Eqs.~\eqref{eq:Psi2evol}--\eqref{eq:E2evol} are solved for $\Psi_2$ and $\hat E_2$, with no independent second-order homogeneous mode, the Hamiltonian and
momentum constraints reconstruct the matter variables rather than evolve  them,
\begin{equation}
 \delta_2=\frac{S_{00}-6\Hc\dot\Psi_2-2k_L^2\Sigma_2}{3\Hc^2},
 \qquad \Sigma_2\equiv\Psi_2+\frac13\hat E_2 ,
 \label{eq:delta2rec}
\end{equation}
\begin{equation}
 \hat v_2=-\frac{1}{4\Hc^2}\left(2\dot\Sigma_2+\frac{k_is_i}{k_L^2}\right),
 \qquad S_{0i}\equiv is_i .
 \label{eq:v2rec}
\end{equation}
For the full unequal-leg source $k_is_i=\Order(k_L^2)$, so the apparent
$1/k_L^2$ in Eq.~\eqref{eq:v2rec} has a regular strict-soft limit.  This completes the second-order synchronous system. 

\subsection{Pullback to comoving slicing and threading}
\label{app:synch-pullback}

To compare with Sec.~\ref{sec:radiation}, the synchronous solution must be mapped to
the comoving slicing. 
At second order this change of slicing and threading is nonlinear: it contains a second-order generator as well as quadratic actions of the first-order generators on the metric and fluid variables, following the standard gauge-transformation formalism of Malik and Wands \cite{Malik:2008im}.

 An irrotational perfect fluid admits slices orthogonal to
$u_\mu$; writing such a slice as
$\eta=\tau+T(\bm x)$, the comoving condition is
$u_i^{\rm com}=u_i+u_0\partial_iT=0$, and hence
\begin{equation}
 \partial_iT=-\frac{u_i}{u_0} ,
 \label{eq:Tnormal}
\end{equation}
which in synchronous gauge, through quadratic order, is
$-u_i/u_0=\partial_i\hat v+2C_{ij}\partial_j\hat v$.  At first order this gives
$T^{(1)}=\hat v^{(1)}$.  At
second order, removing the common Fourier factor $i$, the quadratic covector
kernel is $W_i=k_{Li}\hat v_2+2\big(C^q_{ij}p_j\hat v_p+C^p_{ij}q_j\hat v_q\big)$,
where the first term is the second-order velocity and the remaining terms are the
first-order metric acting on the first-order velocity.  Its longitudinal part fixes
the second-order time shift,
\begin{equation}
 T_2=\frac{k_{Li}W_i}{k_L^2} .
 \label{eq:T2}
\end{equation}
At finite $\e$, scalar--scalar coupling can also generate a transverse remainder of
$W_i$ that no scalar time shift removes.  Thus ``comoving'' here means
scalar-comoving with isotropic threading, which is the same target gauge used in the
direct calculation.

Completing the isotropic spatial threading is stated directly as a coordinate
map, $x^\mu\to y^\mu$ with synchronous coordinates $x^\mu$ and final
scalar-comoving coordinates $y^\mu$,
\begin{align}
 x^0(y)={}&\eta+A\,T_qe^{i\bm q\cdot\bm y}+B\,T_pe^{i\bm p\cdot\bm y}
 +AB\,T_2e^{i\bm k_L\cdot\bm y} , \nonumber\\
 x^i(y)={}&y^i+iA\,s_q^ie^{i\bm q\cdot\bm y}+iB\,s_p^ie^{i\bm p\cdot\bm y}
 +iAB\,s_2^ie^{i\bm k_L\cdot\bm y} ,
 \label{eq:directmap}
\end{align}
with the second-order spatial pieces $s_2^i$ fixed, alongside $T_2$ above, by the
target gauge conditions.  
\begin{equation}
 k_L^i\,(u_i)^{(2)}_{\rm com}=0 ,\qquad \hat E^{(2)}_{\rm com}=0 ,
 \qquad F_i^{(2)}=0 .
 \label{eq:targetgauge}
\end{equation}
Here $F_i$ denotes the transverse vector part of the
spatial metric, $\partial^iF_i=0$, whose contribution is proportional to
$\partial_iF_j+\partial_jF_i$.
The first condition fixes the scalar-comoving slicing, the second fixes the scalar
spatial threading, and the third removes the vector part of the spatial metric.
These conditions do not license setting $k_L=0$ in advance.  With
$J^\mu{}_\nu=\partial x^\mu/\partial y^\nu$, every field is produced by the same
one map,
\begin{equation}
 g^{\rm com}_{\mu\nu}=J^\alpha{}_\mu J^\beta{}_\nu\,g^{\rm Syn}_{\alpha\beta}(x(y)),
 \qquad u^{\rm com}_\mu=J^\alpha{}_\mu\,u^{\rm Syn}_\alpha(x(y)),
 \qquad \rho_{\rm com}=\rho_{\rm Syn}(x(y)) ,
 \label{eq:onemappullback}
\end{equation}
and, after the target conditions remove the scalar and vector parts of the
transformed spatial metric,
\begin{equation}
 \R_2=\frac16\,h^{{\rm com}(2)}_{ii} .
 \label{eq:Rcomdef}
\end{equation}

\begin{figure}[!htbp]
\centering
\includegraphics[width=0.72\linewidth]{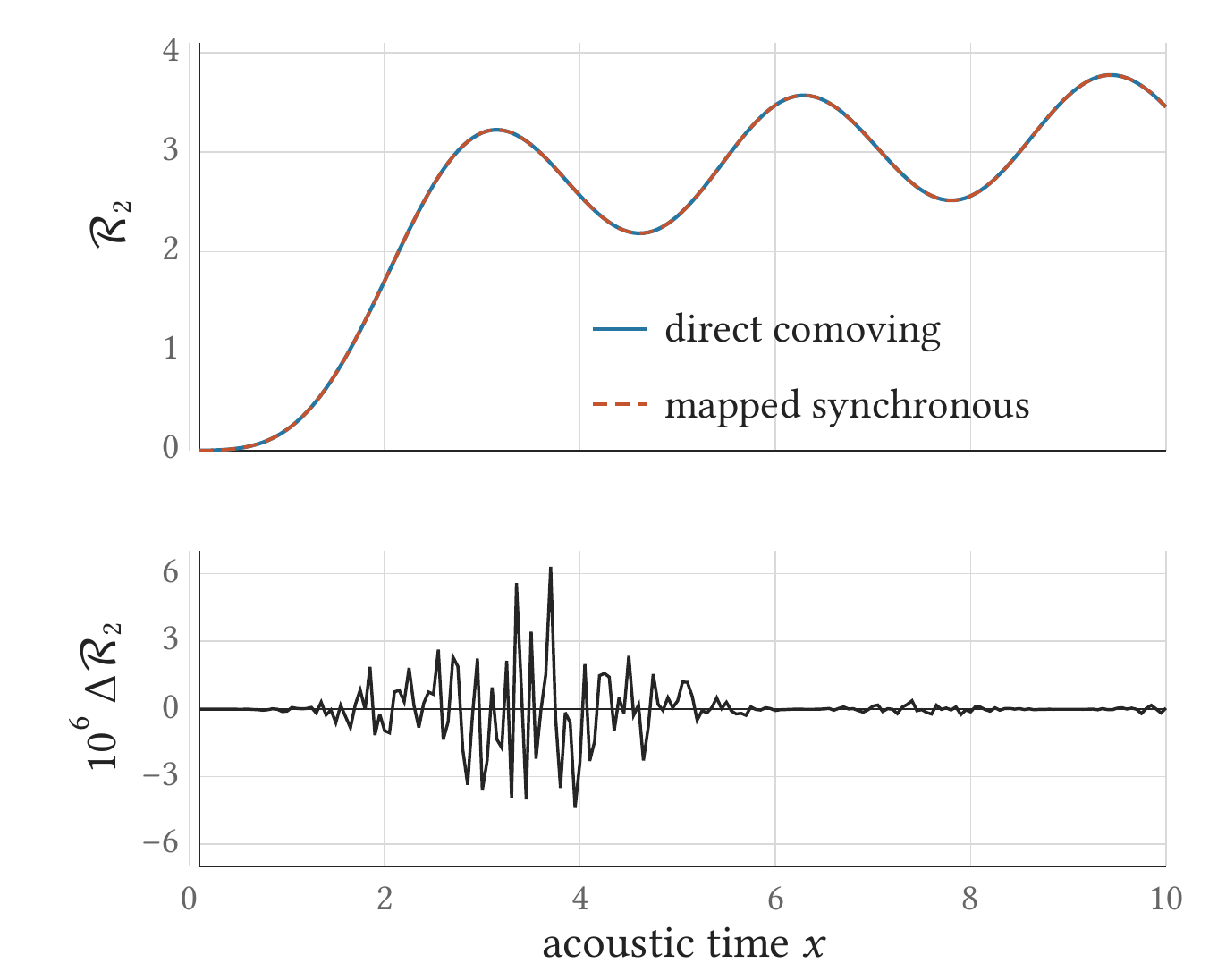}
\caption{Direct comoving-gauge curvature response (solid) compared with the
synchronous-evolution-plus-pullback result (dashed),
at $\e=0.005$ and $\mu=0.5$.  The soft mode stays superhorizon throughout the
range shown.  \emph{Top:} $\R_2$ from both routes, overlaid and visually
indistinguishable at this scale.  \emph{Bottom:} the \emph{signed} difference
$\Delta\R_2=\R_{2,\rm direct}-\R_{2,\rm map}$  in units of $10^{-6}$, consistent with the computational precision of their respective numerical evaluations.}
\label{fig:synch}
\end{figure}

After the pullback, the synchronous construction and the direct calculation are
two representations of the same comoving response.    Figure~\ref{fig:synch}
illustrates their excellent agreement at finite soft momentum.
Both representations approach the regular strict-soft comoving response:  no inverse powers of $k_L$ occur in $\R_2$.
\FloatBarrier

\section{Langlois--Vernizzi conservation and the hard-hard response}
\label{app:lv}

At first sight, the nonzero, oscillating $\R_2$ sourced by hard acoustic legs
(Sec.~\ref{sec:radiation}) appears to conflict with the exact
Langlois--Vernizzi (LV) conservation law \cite{Langlois:2005ii,Langlois:2005qp} for the curvature one-form, which directly applies to the barotropic radiation fluid here.   Here we show that there is no conflict between these statements.

 For a barotropic perfect fluid define
\begin{equation}
 S_{\rm LV}=\alpha_{\rm LV}+\frac14\ln\rho ,
 \qquad u^a\nabla_a\alpha_{\rm LV}=\frac\theta3 ,
 \label{eq:SLV}
\end{equation}
Here $\theta\equiv\nabla_a u^a$ is the covariant expansion of the fluid congruence.
In the scalar comoving sector, where the hypersurface normal is $n^a=u^a$, it is
minus the ADM trace, $K=-\theta$, in the convention of Sec.~\ref{sec:ham}; away from that
identification $K$ is slice dependent, whereas $\theta$ remains the fluid expansion.  Thus
$\alpha_{\rm LV}$ integrates the local expansion history along a fluid worldline
--- a clock built from how much a fluid element has expanded --- and $S_{\rm LV}$ combines that clock with density in the one combination that
makes it behave like a curvature potential for a barotropic fluid.  Its spatial
gradient on the hypersurface orthogonal to the fluid,
$\zeta_a=h_a{}^b\nabla_bS_{\rm LV}$ with $h_a{}^b=\delta_a{}^b+u_au^b$, obeys the
exact identity
\begin{equation}
 u^a\nabla_aS_{\rm LV}=\frac{E_\rho}{4\rho} ,
 \qquad E_\rho\equiv u^a\nabla_a\rho+\frac43\theta\rho ,
 \label{eq:LV-radiation-identity}
\end{equation}
where $E_\rho$ is the local violation of the radiation continuity equation.  For an
ideal fluid with $\bar P=w\bar\rho$, $w$ constant, and no entropy perturbation,
$E_\rho=0$ identically.  Thus $\zeta_a$ is \emph{exactly} Lie-transported along the
fluid congruence.  The result holds exactly: no gradient
expansion enters and so it is not restricted to superhorizon scales like other proofs of curvature conservation.

There is no conflict between this exact conservation law and the oscillations and secular growth of $\R_2$.
Lie-transported along $u^a$ means constant
when evaluated at a fixed fluid element, i.e.\ at fixed Lagrangian label $X^I$, not
at a fixed Eulerian coordinate or a fixed Fourier mode.  $\R_2$, by
contrast, is read off as the coefficient of $e^{i\bm k_L\cdot\bm y}$ on a slice at
fixed coordinate $\bm y$.  These agree only if the coordinates $\bm y$ used to state
$\R_2$ are themselves the fixed fluid labels --- and they are not:
although $u_i=0$ in this gauge, the scalar coordinate velocity is $v=-B$ and need not vanish.
``Comoving'' here means \emph{scalar-comoving and isotropically threaded}
(Appendix~\ref{app:synch}): the scalar condition $u_i=0$ fixes the time slicing to
be orthogonal to $u^a$, exactly as $\zeta_a$'s own projector $h_a{}^b$ requires, but
an independent condition,
$\hat E=0$, is imposed to fix the \emph{spatial} threading by isotropy rather than by
following fluid elements.  Maintaining $\hat E=0$ order by order requires the extra
spatial reshuffling recorded in Appendix~\ref{app:synch}'s pullback --- the piece
$s_2^i$, fixed by the target gauge conditions of Eq.~\eqref{eq:targetgauge} --- and,
at finite external momentum, leaves a genuine transverse remainder that no scalar
coordinate choice can absorb (Sec.~\ref{app:synch-pullback}).  Neither piece is a
statement about fluid labels.  So $\bm y\ne X^I$ beyond leading order, by
construction, independent of any dynamics: the isotropizing threading choice, not a
failure of conservation, is what separates the two.

For the finite-$\epsilon$ hard-hard solution of the main text, an independent evaluation of the
continuity residual verifies $E_\rho=0$ through second order.  This explicitly
checks that the solution realizes the LV transport law.  If $X^I$ are fluid
labels, $u^a\partial_aX^I=0$, the pulled-back covector
$\zeta_I\equiv\zeta_a\,\partial x^a/\partial X^I$ consequently obeys
\begin{equation}
 \frac{d\zeta_I}{d\tau}
 =\frac{\partial x^a}{\partial X^I}(\mathcal L_u\zeta)_a
 =\partial_I\!\left(\frac{E_\rho}{4\rho}\right)=0.
 \label{eq:LV-label-transport}
\end{equation}
Thus LV conservation constrains $\zeta_I$, not the Eulerian Fourier coefficient
$\R_2$.  A nonzero, oscillating $\R_2$ is therefore not in tension with exact
LV conservation.  It is only on superhorizon scales for linear modes that the
two coincide.  Here the hard modes have already undergone acoustic oscillations
and the two relate to different quantities.

\bibliography{LSWN_references}

%apsrev4-2.bst 2019-01-14 (MD) hand-edited version of apsrev4-1.bst
%Control: key (0)
%Control: author (8) initials jnrlst
%Control: editor formatted (1) identically to author
%Control: production of article title (0) allowed
%Control: page (0) single
%Control: year (1) truncated
%Control: production of eprint (0) enabled
\begin{thebibliography}{13}%
\makeatletter
\providecommand \@ifxundefined [1]{%
 \@ifx{#1\undefined}
}%
\providecommand \@ifnum [1]{%
 \ifnum #1\expandafter \@firstoftwo
 \else \expandafter \@secondoftwo
 \fi
}%
\providecommand \@ifx [1]{%
 \ifx #1\expandafter \@firstoftwo
 \else \expandafter \@secondoftwo
 \fi
}%
\providecommand \natexlab [1]{#1}%
\providecommand \enquote  [1]{``#1''}%
\providecommand \bibnamefont  [1]{#1}%
\providecommand \bibfnamefont [1]{#1}%
\providecommand \citenamefont [1]{#1}%
\providecommand \href@noop [0]{\@secondoftwo}%
\providecommand \href [0]{\begingroup \@sanitize@url \@href}%
\providecommand \@href[1]{\@@startlink{#1}\@@href}%
\providecommand \@@href[1]{\endgroup#1\@@endlink}%
\providecommand \@sanitize@url [0]{\catcode `\\12\catcode `\$12\catcode
  `\&12\catcode `\#12\catcode `\^12\catcode `\_12\catcode `\%12\relax}%
\providecommand \@@startlink[1]{}%
\providecommand \@@endlink[0]{}%
\providecommand \url  [0]{\begingroup\@sanitize@url \@url }%
\providecommand \@url [1]{\endgroup\@href {#1}{\urlprefix }}%
\providecommand \urlprefix  [0]{URL }%
\providecommand \Eprint [0]{\href }%
\providecommand \doibase [0]{https://doi.org/}%
\providecommand \selectlanguage [0]{\@gobble}%
\providecommand \bibinfo  [0]{\@secondoftwo}%
\providecommand \bibfield  [0]{\@secondoftwo}%
\providecommand \translation [1]{[#1]}%
\providecommand \BibitemOpen [0]{}%
\providecommand \bibitemStop [0]{}%
\providecommand \bibitemNoStop [0]{.\EOS\space}%
\providecommand \EOS [0]{\spacefactor3000\relax}%
\providecommand \BibitemShut  [1]{\csname bibitem#1\endcsname}%
\let\auto@bib@innerbib\@empty
%</preamble>
\bibitem [{\citenamefont {Barenboim}\ \emph
  {et~al.}(2025{\natexlab{a}})\citenamefont {Barenboim}, \citenamefont
  {Ireland},\ and\ \citenamefont {Stebbins}}]{Barenboim:2025ccc}%
  \BibitemOpen
  \bibfield  {author} {\bibinfo {author} {\bibfnamefont {G.}~\bibnamefont
  {Barenboim}}, \bibinfo {author} {\bibfnamefont {A.}~\bibnamefont {Ireland}},\
  and\ \bibinfo {author} {\bibfnamefont {A.}~\bibnamefont {Stebbins}},\
  }\bibfield  {title} {\bibinfo {title} {{Large Scale White Noise and
  Cosmology}},\ }\href@noop {} {\  (\bibinfo {year} {2025}{\natexlab{a}})},\
  \Eprint {https://arxiv.org/abs/2511.13866} {arXiv:2511.13866 [astro-ph.CO]}
  \BibitemShut {NoStop}%
\bibitem [{\citenamefont {Barenboim}\ \emph
  {et~al.}(2025{\natexlab{b}})\citenamefont {Barenboim}, \citenamefont
  {Ireland},\ and\ \citenamefont {Stebbins}}]{Barenboim:2025jdg}%
  \BibitemOpen
  \bibfield  {author} {\bibinfo {author} {\bibfnamefont {G.}~\bibnamefont
  {Barenboim}}, \bibinfo {author} {\bibfnamefont {A.}~\bibnamefont {Ireland}},\
  and\ \bibinfo {author} {\bibfnamefont {A.}~\bibnamefont {Stebbins}},\
  }\bibfield  {title} {\bibinfo {title} {{The Noisy Universe}},\ }\href@noop {}
  {\  (\bibinfo {year} {2025}{\natexlab{b}})},\ \Eprint
  {https://arxiv.org/abs/2511.15803} {arXiv:2511.15803 [astro-ph.CO]}
  \BibitemShut {NoStop}%
\bibitem [{\citenamefont {Langlois}\ and\ \citenamefont
  {Vernizzi}(2005{\natexlab{a}})}]{Langlois:2005qp}%
  \BibitemOpen
  \bibfield  {author} {\bibinfo {author} {\bibfnamefont {D.}~\bibnamefont
  {Langlois}}\ and\ \bibinfo {author} {\bibfnamefont {F.}~\bibnamefont
  {Vernizzi}},\ }\bibfield  {title} {\bibinfo {title} {{Conserved non-linear
  quantities in cosmology}},\ }\href
  {https://doi.org/10.1103/PhysRevD.72.103501} {\bibfield  {journal} {\bibinfo
  {journal} {Phys. Rev. D}\ }\textbf {\bibinfo {volume} {72}},\ \bibinfo
  {pages} {103501} (\bibinfo {year} {2005}{\natexlab{a}})},\ \Eprint
  {https://arxiv.org/abs/astro-ph/0509078} {arXiv:astro-ph/0509078}
  \BibitemShut {NoStop}%
\bibitem [{\citenamefont {Stebbins}(2023)}]{Stebbins:2023gic}%
  \BibitemOpen
  \bibfield  {author} {\bibinfo {author} {\bibfnamefont {A.}~\bibnamefont
  {Stebbins}},\ }\bibfield  {title} {\bibinfo {title} {{Generation of
  Isocurvature from Curvature Inhomogeneities on Super-Horizon Scales}},\
  }\href@noop {} {\  (\bibinfo {year} {2023})},\ \Eprint
  {https://arxiv.org/abs/2311.17379} {arXiv:2311.17379 [astro-ph.CO]}
  \BibitemShut {NoStop}%
\bibitem [{\citenamefont {Stebbins}(2026)}]{Stebbins:2026stf}%
  \BibitemOpen
  \bibfield  {author} {\bibinfo {author} {\bibfnamefont {A.}~\bibnamefont
  {Stebbins}},\ }\bibfield  {title} {\bibinfo {title} {{A Space-Time Fluid
  (Unabridged)}},\ }\href@noop {} {\  (\bibinfo {year} {2026})},\ \Eprint
  {https://arxiv.org/abs/2601.16996} {arXiv:2601.16996 [physics.gen-ph]}
  \BibitemShut {NoStop}%
\bibitem [{\citenamefont {Tsagas}\ \emph {et~al.}(2008)\citenamefont {Tsagas},
  \citenamefont {Challinor},\ and\ \citenamefont {Maartens}}]{Tsagas:2007yx}%
  \BibitemOpen
  \bibfield  {author} {\bibinfo {author} {\bibfnamefont {C.~G.}\ \bibnamefont
  {Tsagas}}, \bibinfo {author} {\bibfnamefont {A.}~\bibnamefont {Challinor}},\
  and\ \bibinfo {author} {\bibfnamefont {R.}~\bibnamefont {Maartens}},\
  }\bibfield  {title} {\bibinfo {title} {{Relativistic cosmology and
  large-scale structure}},\ }\href
  {https://doi.org/10.1016/j.physrep.2008.03.003} {\bibfield  {journal}
  {\bibinfo  {journal} {Phys. Rept.}\ }\textbf {\bibinfo {volume} {465}},\
  \bibinfo {pages} {61} (\bibinfo {year} {2008})},\ \Eprint
  {https://arxiv.org/abs/0705.4397} {arXiv:0705.4397 [astro-ph]} \BibitemShut
  {NoStop}%
\bibitem [{\citenamefont {Inomata}\ \emph {et~al.}(2023)\citenamefont
  {Inomata}, \citenamefont {Lee},\ and\ \citenamefont {Hu}}]{Inomata:2023faq}%
  \BibitemOpen
  \bibfield  {author} {\bibinfo {author} {\bibfnamefont {K.}~\bibnamefont
  {Inomata}}, \bibinfo {author} {\bibfnamefont {H.}~\bibnamefont {Lee}},\ and\
  \bibinfo {author} {\bibfnamefont {W.}~\bibnamefont {Hu}},\ }\bibfield
  {title} {\bibinfo {title} {{Synchronizing the consistency relation}},\ }\href
  {https://doi.org/10.1088/1475-7516/2023/08/021} {\bibfield  {journal}
  {\bibinfo  {journal} {JCAP}\ }\textbf {\bibinfo {volume} {08}},\ \bibinfo
  {pages} {021}},\ \Eprint {https://arxiv.org/abs/2304.10559} {arXiv:2304.10559
  [astro-ph.CO]} \BibitemShut {NoStop}%
\bibitem [{\citenamefont {Zeldovich}(1970)}]{Zeldovich:1969sb}%
  \BibitemOpen
  \bibfield  {author} {\bibinfo {author} {\bibfnamefont {Y.~B.}\ \bibnamefont
  {Zeldovich}},\ }\bibfield  {title} {\bibinfo {title} {{Gravitational
  instability: An Approximate theory for large density perturbations}},\
  }\href@noop {} {\bibfield  {journal} {\bibinfo  {journal} {Astron.
  Astrophys.}\ }\textbf {\bibinfo {volume} {5}},\ \bibinfo {pages} {84}
  (\bibinfo {year} {1970})}\BibitemShut {NoStop}%
\bibitem [{\citenamefont {Peebles}(1980)}]{Peebles:1980yev}%
  \BibitemOpen
  \bibfield  {author} {\bibinfo {author} {\bibfnamefont {P.~J.}\ \bibnamefont
  {Peebles}},\ }\href@noop {} {\emph {\bibinfo {title} {{The Large-Scale
  Structure of the Universe}}}}\ (\bibinfo  {publisher} {Princeton University
  Press},\ \bibinfo {year} {1980})\BibitemShut {NoStop}%
\bibitem [{\citenamefont {Bartolo}\ \emph {et~al.}(2005)\citenamefont
  {Bartolo}, \citenamefont {Matarrese},\ and\ \citenamefont
  {Riotto}}]{Bartolo:2005xa}%
  \BibitemOpen
  \bibfield  {author} {\bibinfo {author} {\bibfnamefont {N.}~\bibnamefont
  {Bartolo}}, \bibinfo {author} {\bibfnamefont {S.}~\bibnamefont {Matarrese}},\
  and\ \bibinfo {author} {\bibfnamefont {A.}~\bibnamefont {Riotto}},\
  }\bibfield  {title} {\bibinfo {title} {{Signatures of primordial
  non-Gaussianity in the large-scale structure of the Universe}},\ }\href
  {https://doi.org/10.1088/1475-7516/2005/10/010} {\bibfield  {journal}
  {\bibinfo  {journal} {JCAP}\ }\textbf {\bibinfo {volume} {10}},\ \bibinfo
  {pages} {010}},\ \Eprint {https://arxiv.org/abs/astro-ph/0501614}
  {arXiv:astro-ph/0501614} \BibitemShut {NoStop}%
\bibitem [{\citenamefont {Green}\ and\ \citenamefont
  {Wald}(2011)}]{Green:2010qy}%
  \BibitemOpen
  \bibfield  {author} {\bibinfo {author} {\bibfnamefont {S.~R.}\ \bibnamefont
  {Green}}\ and\ \bibinfo {author} {\bibfnamefont {R.~M.}\ \bibnamefont
  {Wald}},\ }\bibfield  {title} {\bibinfo {title} {{A new framework for
  analyzing the effects of small scale inhomogeneities in cosmology}},\ }\href
  {https://doi.org/10.1103/PhysRevD.83.084020} {\bibfield  {journal} {\bibinfo
  {journal} {Phys. Rev. D}\ }\textbf {\bibinfo {volume} {83}},\ \bibinfo
  {pages} {084020} (\bibinfo {year} {2011})},\ \Eprint
  {https://arxiv.org/abs/1011.4920} {arXiv:1011.4920 [gr-qc]} \BibitemShut
  {NoStop}%
\bibitem [{\citenamefont {Malik}\ and\ \citenamefont
  {Wands}(2009)}]{Malik:2008im}%
  \BibitemOpen
  \bibfield  {author} {\bibinfo {author} {\bibfnamefont {K.~A.}\ \bibnamefont
  {Malik}}\ and\ \bibinfo {author} {\bibfnamefont {D.}~\bibnamefont {Wands}},\
  }\bibfield  {title} {\bibinfo {title} {{Cosmological perturbations}},\ }\href
  {https://doi.org/10.1016/j.physrep.2009.03.001} {\bibfield  {journal}
  {\bibinfo  {journal} {Phys. Rept.}\ }\textbf {\bibinfo {volume} {475}},\
  \bibinfo {pages} {1} (\bibinfo {year} {2009})},\ \Eprint
  {https://arxiv.org/abs/0809.4944} {arXiv:0809.4944 [astro-ph]} \BibitemShut
  {NoStop}%
\bibitem [{\citenamefont {Langlois}\ and\ \citenamefont
  {Vernizzi}(2005{\natexlab{b}})}]{Langlois:2005ii}%
  \BibitemOpen
  \bibfield  {author} {\bibinfo {author} {\bibfnamefont {D.}~\bibnamefont
  {Langlois}}\ and\ \bibinfo {author} {\bibfnamefont {F.}~\bibnamefont
  {Vernizzi}},\ }\bibfield  {title} {\bibinfo {title} {{Evolution of non-linear
  cosmological perturbations}},\ }\href
  {https://doi.org/10.1103/PhysRevLett.95.091303} {\bibfield  {journal}
  {\bibinfo  {journal} {Phys. Rev. Lett.}\ }\textbf {\bibinfo {volume} {95}},\
  \bibinfo {pages} {091303} (\bibinfo {year} {2005}{\natexlab{b}})},\ \Eprint
  {https://arxiv.org/abs/astro-ph/0503416} {arXiv:astro-ph/0503416}
  \BibitemShut {NoStop}%
\end{thebibliography}%
\end{document}